%% file: main.tex
\documentclass[runningheads,table]{llncs}
\usepackage{pgfplotstable}
\usepackage{booktabs}
\usepackage{tabularx}
\usepackage[T1]{fontenc}
\usepackage{graphicx}
\usepackage{multirow}
\usepackage{caption}
\usepackage{booktabs}
\usepackage{tikz,hyperref}
\usepackage{subcaption}
\usepackage{adjustbox}
\usepackage{rotating}
\usepackage{listings}
\usepackage{pgfplots}
\pgfplotsset{compat=1.18}
\usepackage{pifont}
\usepackage{todonotes}
\definecolor{col1}{HTML}{F67280}

\definecolor{dkgreen}{rgb}{0,0.6,0}
\definecolor{gray}{rgb}{0.5,0.5,0.5}
\definecolor{mauve}{rgb}{0.58,0,0.82}

\definecolor{lime}{HTML}{A6CE39}
\DeclareRobustCommand{\orcidicon}{
	\begin{tikzpicture}
	\draw[lime, fill=lime] (0,0) 
	circle [radius=0.16] 
	node[white] {{\fontfamily{qag}\selectfont \tiny ID}};
	\draw[white, fill=white] (-0.0625,0.095) 
	circle [radius=0.007];
	\end{tikzpicture}
	\hspace{-2mm}
}
\foreach \x in {A, ..., Z}{\expandafter\xdef\csname orcid\x\endcsname{\noexpand\href{https://orcid.org/\csname orcidauthor\x\endcsname}
			{\noexpand\orcidicon}}
}

\begin{document}
\fontsize{10pt}{12pt}\selectfont
\title{Evaluating Pre-Trained Models for Multi-Language Vulnerability Patching}
\titlerunning{}

\author{
Zanis Ali Khan\orcidA
\and
Aayush Garg\orcidB
\and
Yuejun Guo\orcidC
\and
Qiang Tang\orcidD}

\authorrunning{Khan et al.}

\institute{Luxembourg Institute of Science and Technology (LIST), Luxembourg \\
\email{}
}

\maketitle 

\begin{abstract}

Software vulnerabilities pose critical security risks, demanding prompt and effective mitigation strategies. While advancements in Automated Program Repair (APR) have primarily targeted general software bugs, the domain of vulnerability patching, which is a security-critical subset of APR, remains underexplored. This paper investigates the potential of pre-trained language models, CodeBERT and CodeT5, for automated vulnerability patching across diverse datasets and five programming languages. We evaluate these models on their accuracy, computational efficiency, and how the length of vulnerable code patches impacts performance. Our findings reveal promising accuracy levels, particularly for CodeT5 on datasets with complex vulnerability patterns, while CodeBERT demonstrates strengths in handling fragmented or context-limited datasets. CodeT5 further showcases superior efficiency, making it well-suited for large-scale applications. However, both models face challenges in maintaining performance as patch length increases, highlighting the complexity of addressing extended in program repair specifically aimed at fixing vulnerabilities. This study benchmarks model performance, highlights key limitations, and offers insights to improve automated vulnerability patching for practical security applications.

\keywords{vulnerability patching \and code patching \and automated program repair}
\end{abstract}

\section{Introduction}\label{sec:introduction}

Software vulnerabilities pose a persistent threat to modern software systems, exposing them to potential exploitation by attackers. These vulnerabilities, ranging from memory management issues to injection flaws, can lead to unauthorized access, data breaches, and service disruptions~\cite{OkutanMMKSGS23}. Addressing these vulnerabilities promptly is critical to maintain the security and reliability of software systems~\cite{AlbaneseCJ14}. However, the manual process of identifying and remediating these issues is labor-intensive, error-prone, and unable to keep up with the increasing complexity and scale of modern software ecosystems\cite{KhanP18}.

Automated Program Repair (APR) has become a promising approach to address this issue, utilizing computational methods to generate bug patches autonomously~\cite{BuiPVMS24}. While APR has seen significant advancements in fixing general software bugs, vulnerability-focused program repair—a domain targeting the unique challenges of security vulnerabilities—remains underexplored. Unlike functional bugs, vulnerabilities require patches that address not only the immediate issue but also ensure that the fix is secure, robust, and resistant to future exploitation~\cite{deFiteroDominguezGGM24}. This added complexity makes vulnerability patching a particularly demanding and challenging subset of APR.

Existing approaches in vulnerability-focused APR rely on either static analysis tools or traditional machine learning models trained on specific vulnerability patterns. While these methods have shown promise in detecting vulnerabilities, their ability to generate meaningful and effective patches is limited. Static analysis tools, for example, excel at identifying vulnerabilities but often fail to produce usable fixes~\cite{PiskachevBB23}. Similarly, conventional machine learning models are constrained by their reliance on limited datasets~\cite{abs-2012-11701}, which hampers their generalizability and effectiveness across diverse programming languages and vulnerability types~\cite{RisseB24}.

Recent progress in deep learning, especially with the advent of pre-trained language models for code like CodeBERT~\cite{feng2020codebert} and CodeT5~\cite{wang2021codet5}, offers a new avenue for automated vulnerability patching. These models leverage large-scale code corpora to learn syntactic and semantic representations of programming languages~\cite{GharibiSF24}, enabling them to perform complex tasks like code generation, summarization, and translation~\cite{GargDPT24}. Their capacity to identify patterns from extensive datasets makes them well-equipped to address the complexities of vulnerability-focused program repair. However, applying these models to vulnerability patching is far from straightforward~\cite{Dang2024}. Programming languages differ significantly in syntax, semantics, and vulnerability patterns, and most pre-trained models are designed for monolingual or narrowly scoped tasks. Consequently, evaluating their effectiveness across multiple programming languages is critical yet underexplored.

In this paper, we address these gaps by systematically evaluating the performance of pre-trained language models in vulnerability-focused program repair across multiple programming languages. Specifically, we investigate the capabilities of CodeBERT and CodeT5 in generating patches for known vulnerabilities across 9 datasets spanning five programming languages. Our evaluation examines the effectiveness of these models in terms of their ability to generate accurate patches and their computational efficiency in handling large-scale vulnerability patching tasks. Additionally, we examined the impact of patch length on the accuracy of \emph{CodeBERT} and \emph{CodeT5} by using \emph{CodeBLEU} and \emph{CrystalBLEU} scores across nine datasets.

Our results reveal that while both models demonstrate notable strengths in generating vulnerability patches, they also exhibit distinct limitations. CodeT5 generally outperforms CodeBERT in accuracy, particularly on datasets with complex or diverse vulnerability patterns. However, both models face challenges in handling datasets with fragmented contexts or sparse data, which hinder their ability to generate accurate fixes. Additionally, our analysis reveals that increasing patch lengths significantly influence the accuracy of both models, with their performance declining to varying degrees depending on dataset characteristics and LLM architecture.

Hence, our contributions in this paper are threefold:
\begin{itemize}
    \item We provide an evaluation of CodeBERT and CodeT5 for vulnerability-focused program repair, covering a diverse set of 9 datasets across multiple programming languages.
    \item We establish benchmarks for model performance in generating vulnerability patches, serving as a foundation for evaluating pre-trained models in dataset-driven vulnerability patching scenarios.
    \item We identify the challenge posed by increasing patch lengths on model performance, offering insights into their implications for automated vulnerability patching.
\end{itemize}

\section{Background and Related Work}\label{sec:related}
\label{sec:back}

\subsection{Software Vulnerability and Mitigation}  
Software vulnerabilities are security flaws or weaknesses in code that can be exploited by attackers~\cite{ogata2018vetting}. For example, a buffer overflow vulnerability occurs when a program attempts to store more data in a buffer than its allocated capacity, causing the excess to overwrite adjacent memory locations and potentially enabling attackers to execute malicious code~\cite{GargPKT24}. The increasing sophistication of such vulnerabilities presents significant challenges in implementing effective mitigation measures.  

While much of the research has focused on vulnerability detection, fewer efforts have addressed patch generation. In detection, traditional static analysis tools have been widely used, though their rule-based nature often limits their ability to capture complex patterns~\cite{static2017andrei}. AI-based approaches have gained prominence for their ability to analyze large code corpora and identify intricate vulnerabilities. Notable models include \emph{CodeBERT}\cite{codebert2020} and \emph{GraphCodeBERT}\cite{GuoRLFT0ZDSFTDC21}, which have shown promise in source code analysis, including applications in vulnerability detection and analysis~\cite{guo2024outside,abs-2303-04247}. Moreover, large language models (LLMs) such as OpenAI’s GPT-4, Meta AI’s Llama2, and Mistral AI’s Mistral have demonstrated effective adaptation to vulnerability detection tasks~\cite{guo2023empirical}.

On the other hand, generating effective patches remains a formidable challenge. Most automated patch generation research focuses on repairing general buggy code rather than addressing vulnerabilities specifically. The following sections will review approaches in this broader area.

\subsection{Traditional Approaches to Code Repair}
Traditional approaches to automated code repair are broadly categorized into heuristic and constraint-based methods~\cite{goues2019automated}. Heuristic methods explore a search space of possible patches, aiming to identify one that satisfies all test cases. To make this process manageable, techniques like transformation schemas are used to generate candidates efficiently~\cite{liu2019tbar}. Approaches such as GenProg~\cite{Le2012genprog} and PAR~\cite{Kim2013automatic} utilize genetic programming to refine the search, while others rely on random or deterministic search strategies to improve efficiency.

Constraint-based methods, on the other hand, leverage symbolic execution~\cite{Cadar2008klee} to guide patch generation. By abstracting program inputs into symbolic values, these techniques explore multiple input paths simultaneously. Tools like SemFix~\cite{semfix2013} and Angelix~\cite{angelix2016} use symbolic execution to infer repair constraints, while methods such as Nopol~\cite{Nopol2017} focus on specific scenarios, like fixing conditional statements.

\subsection{ML-Based Code Repair}  
Machine learning has become a pivotal approach for automating code repair, leveraging models to generate patches for vulnerabilities and bugs. Early methods primarily relied on neural machine translation (NMT) models with encoder-decoder architectures to transform buggy code into patched counterparts. Tools like SequenceR~\cite{SequenceR2021} and CODIT~\cite{Chakraborty2022edit} introduced attention mechanisms to improve focus on relevant input regions during decoding.  

Recent advancements have shifted toward transformer-based architectures, which excel at capturing long-range dependencies and nuanced contextual information. Enabled by attention mechanisms, these models dynamically prioritize relevant code segments, making them highly effective for program repair. Ding \emph{et al.}~\cite{Ding2020metaphor} demonstrated the potential of transformers for program repair, paving the way for widespread adoption.  

Building on transformers, large language models (LLMs) like CodeBERT~\cite{codebert2020} and TFix~\cite{Berabi2021tfix} have become dominant in vulnerability patching due to their pretraining on extensive code corpora. These models have been fine-tuned for targeted repairs, showing promising results across diverse datasets.  

However, vulnerability patching remains uniquely challenging. Unlike general bug repair, it requires highly contextualized and security-specific modifications, demanding models to generalize effectively across complex scenarios. Addressing these challenges necessitates fine-tuning LLMs and developing techniques to enhance their adaptability to diverse datasets and nuanced security requirements.

\section{Research Questions}\label{sec:rqs}

The use of pre-trained language models in automated program repair (APR) has shown 
promising results, particularly for general bugs. However, their effectiveness in 
vulnerability-focused program repair, which demands highly accurate and context-aware 
patches, remains underexplored~\cite{ZhangFZYSC23}. Unlike general-purpose 
bug fixes, vulnerability patches must address specific security threats while aligning 
with language-specific patterns, coding styles, and contextual nuances.

Prominent models like \emph{CodeBERT} and \emph{CodeT5} have excelled in various code-related tasks~\cite{SalzaSGG23,GargDPT24}. Yet, their ability to adapt to 
vulnerability datasets, particularly those spanning diverse programming contexts, is 
uncertain~\cite{GoelGF22}. Understanding how well these models generalize across 
such datasets is critical for advancing their real-world applicability. Hence, we ask:

\begin{itemize}
\item[] \begin{itemize}
            \item [\textbf{RQ1}] \textbf{Accuracy Across Datasets.} How effectively do CodeBERT and CodeT5 generate accurate patches for known vulnerabilities across diverse datasets?
        \end{itemize}
\end{itemize}


Beyond patch accuracy, scalability is crucial for applying pre-trained models in 
vulnerability repair. Modern software systems generate vast amounts of code, demanding 
solutions that are both accurate and efficient~\cite{GargDMCAPT23}. Optimized processes 
have demonstrated effective scaling across larger systems, making computational efficiency particularly important for integration into Continuous Integration and Continuous Deployment (CI/CD) 
pipelines or large-scale vulnerability assessments~\cite{WangJLYX0L22}. Moreover, scalability ensures that these models can handle the dynamic and evolving nature of software development, where new vulnerabilities are discovered continuously. Evaluating inference time and resource usage is essential to assess their practicality in real-world scenarios. Such assessments provide insights into the trade-offs between model accuracy and efficiency, which are critical for deploying these models at scale. Thus, we ask:

\begin{itemize}
\item[] \begin{itemize}
            \item [\textbf{RQ2}] \textbf{Efficiency and Accuracy Trade-offs.} What are the computational trade-offs, in terms of inference efficiency, when using CodeBERT and CodeT5 for large-scale vulnerability patching tasks?
        \end{itemize}
\end{itemize}

Also, research indicates that as the length of generated text increases, LLMs often experience declines in coherence and accuracy. The existing research has shown that these models often struggle to maintain coherent event sequences in longer texts~\cite{GuanMFLDH20} and suffer significant performance degradation in tasks requiring extended contexts~\cite{LiuLHPBPL24}. Therefore, it is crucial to examine whether \emph{CodeBERT} and \emph{CodeT5} maintain their effectiveness in generating accurate patches as patch length increases. Understanding the relationship between patch length (for LLM-generated patches) and model performance will provide insights into their robustness across varying output lengths. Hence, we ask:

\begin{itemize}
\item[] \begin{itemize}
            \item [\textbf{RQ3}] \textbf{Impact of Patch Length.} How does the length of generated patches affect the accuracy of CodeBERT and CodeT5 in vulnerability-focused program repair?
        \end{itemize}
\end{itemize}

\section{Methodology}\label{sec:method}

\begin{figure}[ht]
\centering
\includegraphics[width=0.90\textwidth]{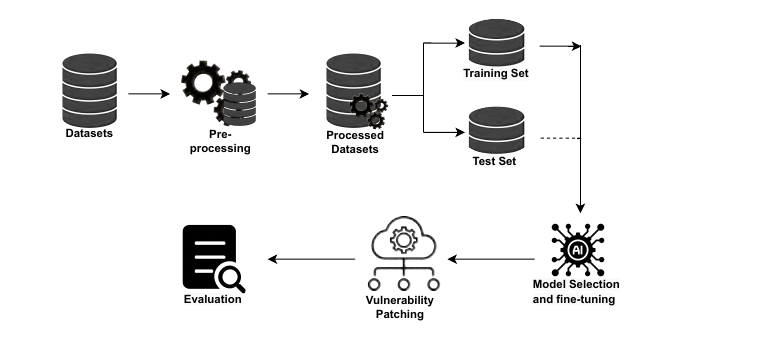}
\caption{Overview of our Methodology.}
\label{methodology}
\end{figure}

The methodology for this study is structured into several stages, as shown in Figure~\ref{methodology}, encompassing dataset collection, preprocessing, training, and evaluation. Below, we describe each stage in detail.

\subsection{Dataset Preparation}
For this study, we collected nine publicly available datasets containing code samples with known vulnerabilities and their corresponding patches. These datasets span multiple programming languages, including Go, PHP, Java, C and C++, ensuring diverse code structures and vulnerability patterns. The inclusion of diverse datasets allowed us to evaluate the models' ability to generalize across varied programming contexts. Details about these datasets, including their references are provided in Section~\ref{sec:datasets}, offering a comprehensive overview of their sources. This diversity in datasets not only enhances the robustness of our evaluation but also reflects real-world scenarios where vulnerabilities span multiple languages and coding paradigms.

\subsection{Preprocessing}\label{subsec:pre-processing}
We preprocessed the raw datasets to standardize their structure and improve compatibility with the models. Given the inherent noise in real-world vulnerability datasets~\cite{abs-2012-11701,KhanSBB24}, our preprocessing aimed to mitigate the effects of noisy or inconsistent data. Accurate preprocessing is crucial for minimizing noise and improving data quality, as emphasized in studies addressing identification challenges in noisy datasets~\cite{Khan0BB22,GargDJCPT22}. By ensuring uniformity and consistency in the datasets, we aimed to create a robust foundation for reliable model training and accurate performance assessment. The following steps performed were critical in reducing noise and preparing the datasets for training and evaluation.
\begin{enumerate}
    \item[i.] \textbf{Token Length Filtering.} Code sequences exceeding \emph{512} tokens were truncated or excluded, as pre-trained models impose this token length limit.
    \item[ii.] \textbf{Comment Removal.} Comments were removed from the code using language-specific regex patterns to focus solely on functional elements of the code.
    \item[iii.] \textbf{Normalization.} Formatting inconsistencies, such as extra whitespace and irregular line breaks, were corrected to ensure uniformity across datasets.
\end{enumerate}

\subsection{Training and Testing Split}
The preprocessed datasets were divided into \emph{85\%} training and \emph{15\%} testing subsets, a common practice in machine learning research to balance model training and evaluation~\cite{0040158}. This split was chosen to maximize training data while preserving sufficient data for evaluation. The separation of training and testing data is crucial for assessing the generalization ability of the model on new, unseen instances. To avoid data leakage, we adhered to best practices for dataset splitting by eliminating any overlapping or duplicate instances between the training and testing sets, ensuring the reliability of experimental outcomes.

\subsection{Model Selection and Fine-Tuning}
We utilized two state-of-the-art pre-trained models, \emph{CodeBERT}\cite{feng2020codebert} and \emph{CodeT5}\cite{wang2021codet5}, to evaluate their performance in vulnerability patching. CodeBERT, designed explicitly for programming tasks, was fine-tuned on the training datasets to learn patterns in vulnerable and patched code. This fine-tuning process allowed CodeBERT to adapt and specialize in recognizing specific vulnerabilities and their corresponding patches. Similarly, CodeT5, a model optimized for code generation and understanding, was fine-tuned to align with the vulnerability patching task. CodeT5’s optimization also focused on enhancing its ability to deal with various code structures and programming languages. The fine-tuning process adjusted the models’ weights based on the training data to optimize their ability to generate accurate patches. While other models like \emph{TFix}\cite{Berabi2021tfix} have shown potential in specific code repair tasks, \emph{CodeBERT} and \emph{CodeT5} are preferred due to their robustness, versatility, and proven effectiveness in real-world applications, offering seamless integration into modern software development workflows.

\subsection{Vulnerability Patching and Evaluation}
During the evaluation phase, the testing datasets were used to assess the models. Each test instance, representing a vulnerable code snippet, was input into the trained models to generate a patched version. These datasets covered a broad spectrum of programming languages and vulnerability types, ensuring a diverse set of conditions for evaluating the models' effectiveness. The patches produced were then compared to the ground truth using the \emph{CodeBLEU}\cite{ren2020codebleu}, and CrystalBLEU~\cite{eghbali2022crystalbleu} metrics, which evaluates the accuracy and quality of the generated patches. By utilizing these metrics, we ensure that our evaluation encompasses both the syntactic correctness and semantics of the patch. CodeBLEU was chosen as it extends the traditional BLEU metric by incorporating code-specific features, such as Abstract Syntax Trees (ASTs) and Semantic Data Flow~\cite{EvtikhievBSB23}. This extension allows CodeBLEU to better capture the structural and logical consistency of code, making it particularly useful for vulnerability patching. While, 
CrystalBLEU incorporates a more nuanced approach by considering trivially shared 
n-grams and offering a refined evaluation metric for code generation tasks.
This ensures that both the syntactic and semantic correctness of the generated patches are assessed, providing a comprehensive measure of model performance. 
Moreover, these metrics together offer a balanced perspective, accounting for both syntactic accuracy and functional correctness, which are essential for high-quality vulnerability repair.

\section{Experimental Setup}\label{sec:eval-subjects}

All experiments in this study were performed on a High-Performance Computing (HPC) cluster featuring nodes equipped with 2.20GHz Intel Xeon Silver 4210 processors and NVIDIA Tesla V100-PCIE-32GB GPUs. Model training and evaluation utilized the PyTorch 2.0.1 framework with CUDA 12 compatibility.

\subsection{Datasets}\label{sec:datasets}

To address the research questions introduced in Section~\ref{sec:rqs}, we utilized publicly 
available datasets containing extensive collections of vulnerable source code and their fixed 
versions as ground truth. Specifically, we employed nine datasets, including 
Go and PHP~\footnote{Go and PHP--\url{https://doi.org/10.5281/zenodo.13870382}}, 
BigVul\footnote{BigVul-- \url{https://github.com/ZeoVan/MSR_20_Code_vulnerability_CSV_Dataset}}\cite{10.1145/3379597.3387501}, 
MegaVul\_Java, MegaVul\_C\_2023, and MegaVul\_C\_2024~\footnote{MegaVul\_Java, MegaVul\_C\_2023, and MegaVul\_C\_2024--\url{https://github.com/Icyrockton/MegaVul}}~\cite{10.1145/3643991.3644886}, SVEN\footnote{SVEN--\url{https://github.com/eth-sri/sven}}\cite{10.1145/3576915.3623175}, and also 
CodeXGlue\_Small\_Java and CodeXGlue\_Medium\_Java~\footnote{CodeXGlue\_Small\_Java and CodeXGlue\_Medium\_Java-- \url{https://github.com/microsoft/CodeXGLUE}}~\cite{lu2021codexgluemachinelearningbenchmark}. These datasets include diverse programming languages such as C, C++, Java, Go, and PHP, providing a comprehensive foundation for 
evaluation. Before using the datasets, we applied preprocessing steps as mentioned in Section~\ref{subsec:pre-processing}.





\begin{table}[htbp] 
\centering 
\small
\caption{Datasets}
\input{tables/datasets}
\label{table:datasets}
\end{table}

Table~\ref{table:datasets} reports on the size of our datasets, in terms of the number of rows 
(\(I_{rows}\)), rows affected by tokenization ($R_{tok.}$), rows affected by 
comment removal ($R_{comm.}$), rows affected by normalization ($R_{norm.}$), 
and the total number of rows remaining after pre-processing ($T_{rows.}$).

\subsection{DL Models}\label{subsec:models}

For vulnerability patching, we considered \emph{CodeBERT}\cite{feng2020codebert} and \emph{CodeT5}\cite{wang2021codet5}, which are widely used for code analysis and vulnerability detection. 

\emph{CodeBERT}\cite{feng2020codebert} is a pre-trained model aimed at bridging the gap between programming and natural languages, improving tasks like code completion, summarization, and vulnerability detection. Using the transformer architecture, it is trained on a diverse dataset of natural language pairs and source code, enabling it to learn both syntactic and semantic relationships. This dual understanding is crucial for accurate vulnerability identification and remediation. Moreover, its scalability makes it an attractive option for projects ranging from small-scale open-source contributions to large enterprise-level software systems.

\emph{CodeT5}\cite{wang2021codet5} is a model optimized for code generation and understanding, excelling in vulnerability detection and patching. Based on the T5 framework, it adapts to programming languages, translating between code and natural language. One of its distinguishing features is its ability to generate context-aware patches that address vulnerabilities while preserving the intent of the original code. Pre-trained on extensive programming data, it generates code, identifies vulnerabilities, and suggests patches with high accuracy. Its ability to handle multiple languages and perform well on benchmarks makes it a powerful tool for software security and code quality improvement. CodeT5 has demonstrated its effectiveness in preserving high-level semantics during decompilation tasks in advancing vulnerability detection and repair  frameworks~\cite{WuJPLD0BS23}.

\subsection{Evaluation Metrics}\label{subsec:metrics}

We used \emph{CrystalBLEU}\cite{eghbali2022crystalbleu} and \emph{CodeBLEU}\cite{ren2020codebleu} as metrics to evaluate the accuracy of the aforementioned LLMs. CrystalBLEU extends BLEU~\cite{papineni2002bleu} by addressing the limitations of traditional n-gram matching, especially when applied to programming languages. Unlike BLEU, which is designed for natural language and fails to capture crucial code-specific syntax and semantics, CrystalBLEU incorporates a more nuanced approach by considering trivially shared n-grams and offering a refined evaluation metric for code generation tasks. Additionally, CodeBLEU improves on BLEU by integrating n-gram matching with abstract syntax tree (AST)-based code structures and semantic analysis through data flow, making it particularly suitable for evaluating the quality of code. Both CrystalBLEU and CodeBLEU provide more accurate and meaningful assessments of code generation models by considering both syntactic and semantic aspects of the generated code.

\section{Results and Discussion}
\label{sec:results}




\subsection{Accuracy Across Datasets (RQ1)}\label{accuracy-comparison}

Table~\ref{table:rq1_results} displays the CodeBLEU, and CrystalBLEU scores of CodeBERT and CodeT5 across nine datasets used in our evaluation. Examining the performance of both models on these datasets reveals key insights into how pre-training data diversity and model architecture impact the models’ effectiveness in vulnerability patching tasks. CodeT5 consistently outperforms CodeBERT across the majority of datasets, with notable improvements in datasets like BigVul, MegaVul\_Java, and SVEN. While the differences are less pronounced on MegaVul\_C\_2023 and MegaVul\_C\_2024, CodeT5 still demonstrates a clear advantage over CodeBERT when evaluated using both CodeBLEU and CrystalBLEU accuracy scores. This result aligns with the fact that CodeT5 has been pre-trained on diverse data that spans a variety of programming languages and textual formats, enabling it to capture more generalized patterns and nuances in code. In particular, the BigVul and MegaVul\_Java datasets—where CodeT5’s performance excels—include a range of real-world vulnerabilities and coding structures that may benefit from CodeT5’s wider pre-training coverage.

\begin{table}[htbp] 
\centering 
\small
\caption{Accuracy Scores}
\input{tables/RQ1_results}

\label{table:rq1_results}
\end{table}

The Go and PHP datasets present exceptions, with CodeBERT achieving higher CodeBLEU scores. By analyzing 
these two datasets, we observed that they often contain incomplete functions or isolated snippets 
lacking full context. This could potentially lead to lower performance for CodeT5, as it relies 
on contextual understanding from diverse sources that might not align well with fragmented or 
incomplete code. Conversely, CodeBERT, which is also trained on a broad variety of programming 
languages, may still benefit from its fine-tuned focus on code structure, making it more adaptable to such fragments.

These findings suggest that CodeBERT's architecture might be inherently more robust when handling 
incomplete or context-limited code, a factor that could contribute to its better performance on Go 
and PHP. Moreover, despite CodeBERT generally being outperformed by CodeT5, its comparable 
performance on CodeXGlue datasets further emphasizes its effectiveness in Java-related tasks. 
This outcome suggests that CodeBERT, though lacking the extensive pre-training diversity of 
CodeT5, can still achieve near-competitive results in certain domains, particularly for language-specific tasks. It is worth noting that the CrystalBLEU score for CodeBERT on 
the SVEN dataset (\textit{0\%}) contrasts sharply with the CodeBLEU score on the same 
dataset (\textit{0.21\%}), highlighting an interesting anomaly specific to SVEN. 
While CodeBLEU has proven effective on other datasets, this discrepancy suggests that 
certain characteristics of the SVEN dataset may influence how these metrics evaluate 
patch quality. A closer manual examination of the patches generated by CodeBERT for 
SVEN revealed that they were often highly buggy and significantly different from the 
ground truth. This observation underscores the need for further investigation to better 
understand the interplay between dataset characteristics and metric sensitivity, rather 
than drawing generalized conclusions about the performance of CodeBLEU or CrystalBLEU.

Our results highlight the benefits of model diversity in deep learning-based vulnerability 
patching. CodeT5's broad pre-training excels on datasets with complex vulnerabilities, 
while CodeBERT's focused design performs well on datasets with more traditional, 
syntactically constrained samples. These insights show that model choice should 
depend on dataset characteristics. CodeBERT's simpler architecture likely makes 
it less reliant on context, while CodeT5 handles diverse inputs more effectively. 
Thus, while CodeT5 is suited for complex, varied data, CodeBERT is valuable in 
environments with incomplete or non-standard code snippets.


\subsection{Efficiency Trade-offs (RQ2)}\label{efficiency-comparison}

Table~\ref{table:rq2_results} presents the average inference efficiency of CodeBERT and CodeT5 across 
nine datasets, measured in execution time per test instance (in seconds). A comparative analysis 
of the results highlights a noticeable performance advantage for CodeT5, which consistently achieves 
faster inference times across different datasets when compared to CodeBERT. This performance difference 
underscores the enhanced efficiency of CodeT5, especially significant given the growing need for 
scalable, real-time processing in software engineering tasks.

The datasets in this analysis vary substantially in test instance counts and domain focus, with smaller, 
dataset like SVEN containing 105 test instances, and larger test sets 
like $CodeXGlue\_Medium\_Java$ encompassing 9,819 instances. It is important to note that 
the column $TestInstances$  in Table~\ref{table:rq2_results} 
reflect only the number of instances present in the test datasets. This diversity in 
test instance quantity and domain specificity helps evaluate the inference efficiency of 
CodeBERT and CodeT5 across varied real-world conditions, from niche, targeted datasets to 
larger, general-purpose code datasets. Despite 
this variability, CodeT5 maintains lower execution times, revealing its robust ability to handle 
diverse dataset characteristics and volumes more effectively. For instance, on $MegaVul\_C\_2024$, 
CodeT5 demonstrates a marked reduction in execution time (1.5246s per instance) compared to 
CodeBERT’s 2.0295s, suggesting a consistent edge in managing complex vulnerability datasets. 
This efficiency gain is even more pronounced in the Java datasets (except $MegaVul\_Java$), 
where CodeT5's execution time is less than half that of CodeBERT, reflecting its adaptability 
and optimized performance for higher data throughput demanding tasks.

\begin{table}[htbp] 
\centering 
\small
\caption{Efficiency Comparison (\textit{In Seconds})}
\input{tables/RQ2_results}
\label{table:rq2_results}
\end{table}

The motivation for focusing on inference efficiency in models like CodeBERT and CodeT5 stems from 
the increasing integration of machine learning models in continuous integration and deployment 
pipelines, where rapid processing is critical. Lower inference time translate directly to cost 
savings in computational resources, making models like CodeT5 more viable for industrial 
applications where high-volume, real-time processing is essential. Moreover, the faster 
execution times associated with CodeT5 enable researchers and practitioners to experiment with 
larger datasets and iterate more frequently, accelerating the model development and testing phases.


In summary, Table~\ref{table:rq2_results} illustrates that CodeT5 not only outperforms CodeBERT in 
average inference time but does so consistently across diverse datasets, making it a promising 
tool for efficient and scalable software engineering solutions. This efficiency advantage may 
guide future choices in model selection for applications requiring low latency and high 
throughput, particularly as the demand for real-time code analysis continues to grow.

\subsection{Impact of Patch Length (RQ3)}

\begin{figure}
  \centering

  \renewcommand{\arraystretch}{1.3}  
    \begin{tabularx}{\textwidth}{
          l X  @{\hspace{3pt}}  X@{\hspace{1pt}}  l 
    }

      & \raisebox{-0.5\height}{\includegraphics[width=0.34\textwidth] {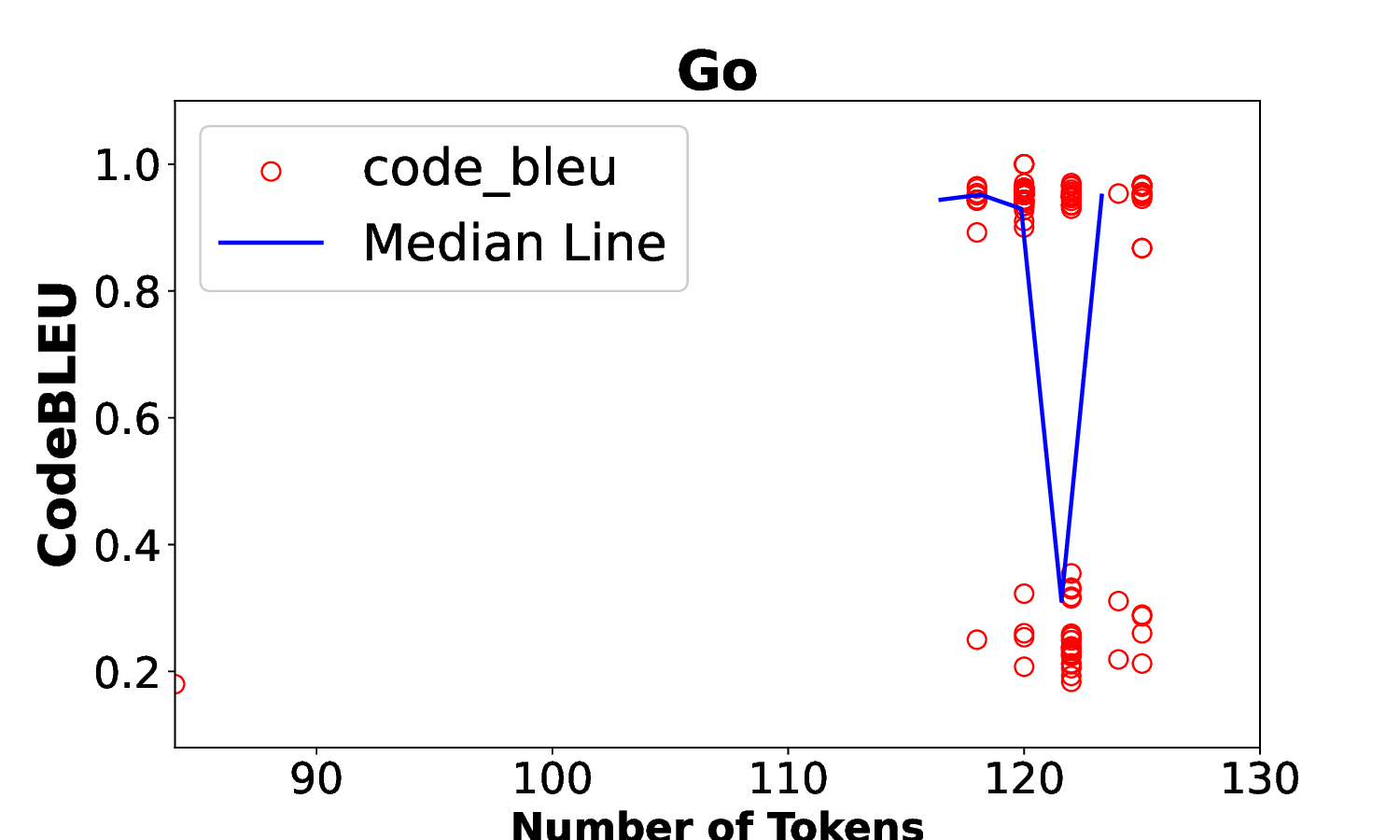}}
      & \raisebox{-0.5\height}{\includegraphics[width=0.34\textwidth] {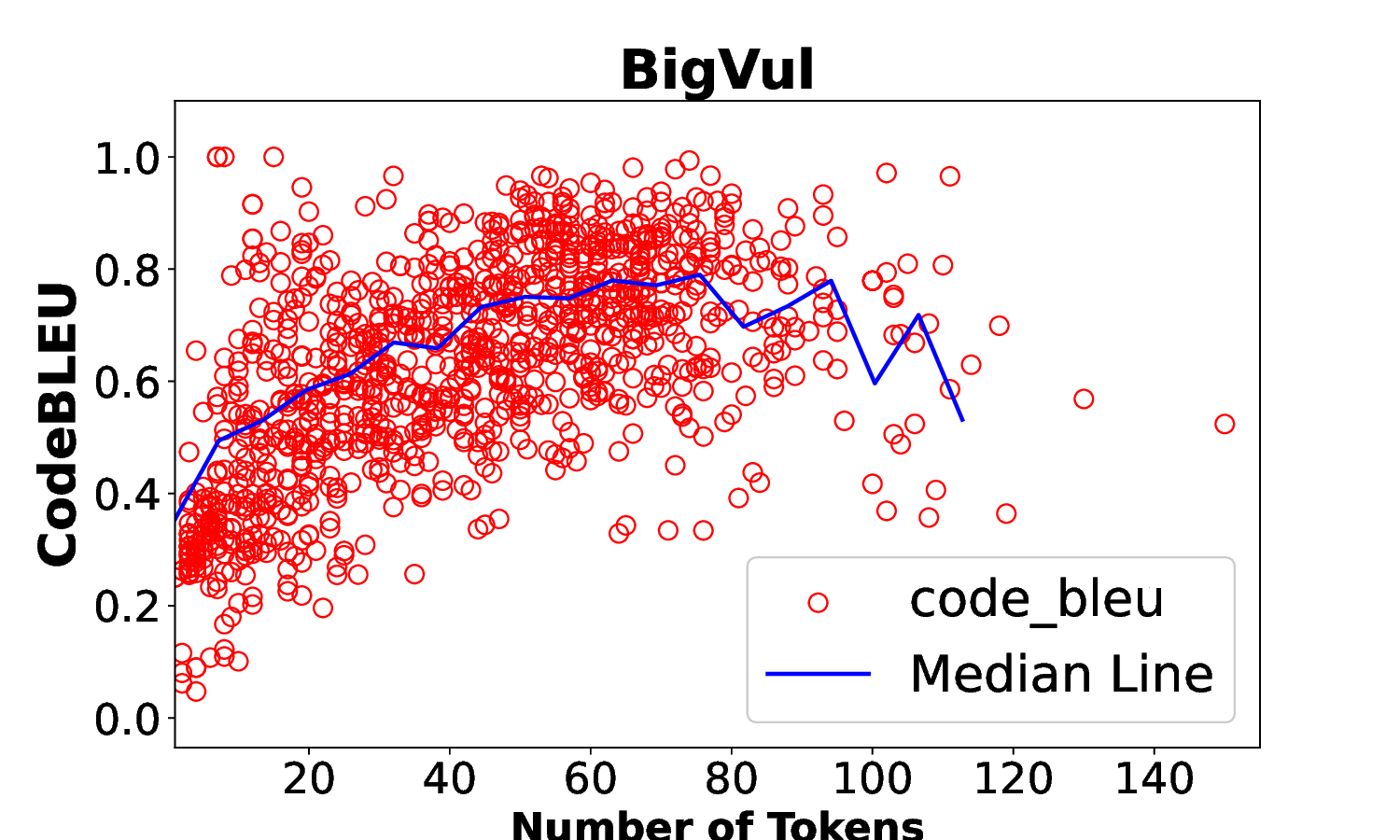}}
       & \raisebox{-0.5\height}{\includegraphics[width=0.34\textwidth] {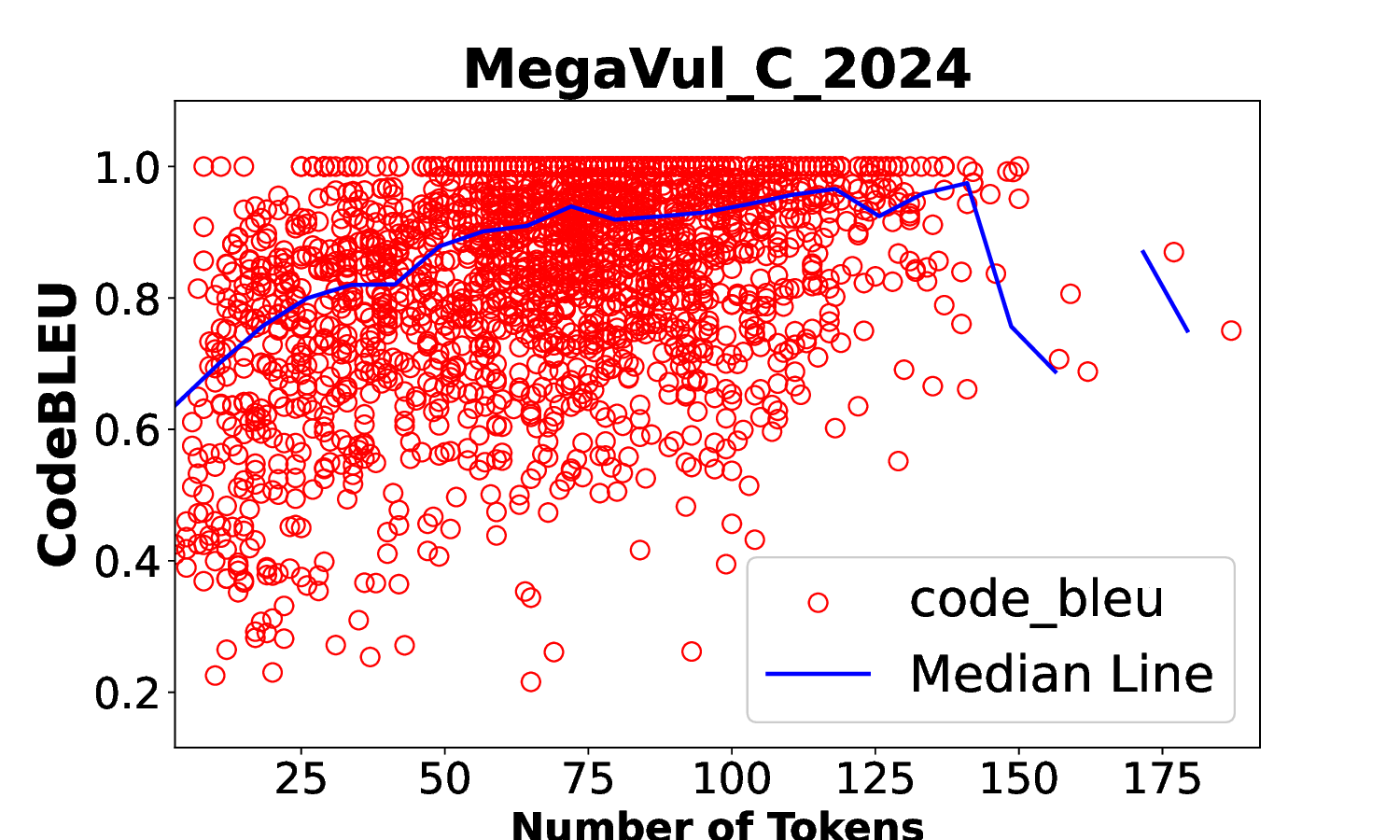}} \\
      
      &  \raisebox{-0.5\height}{\includegraphics[width=0.34\textwidth] {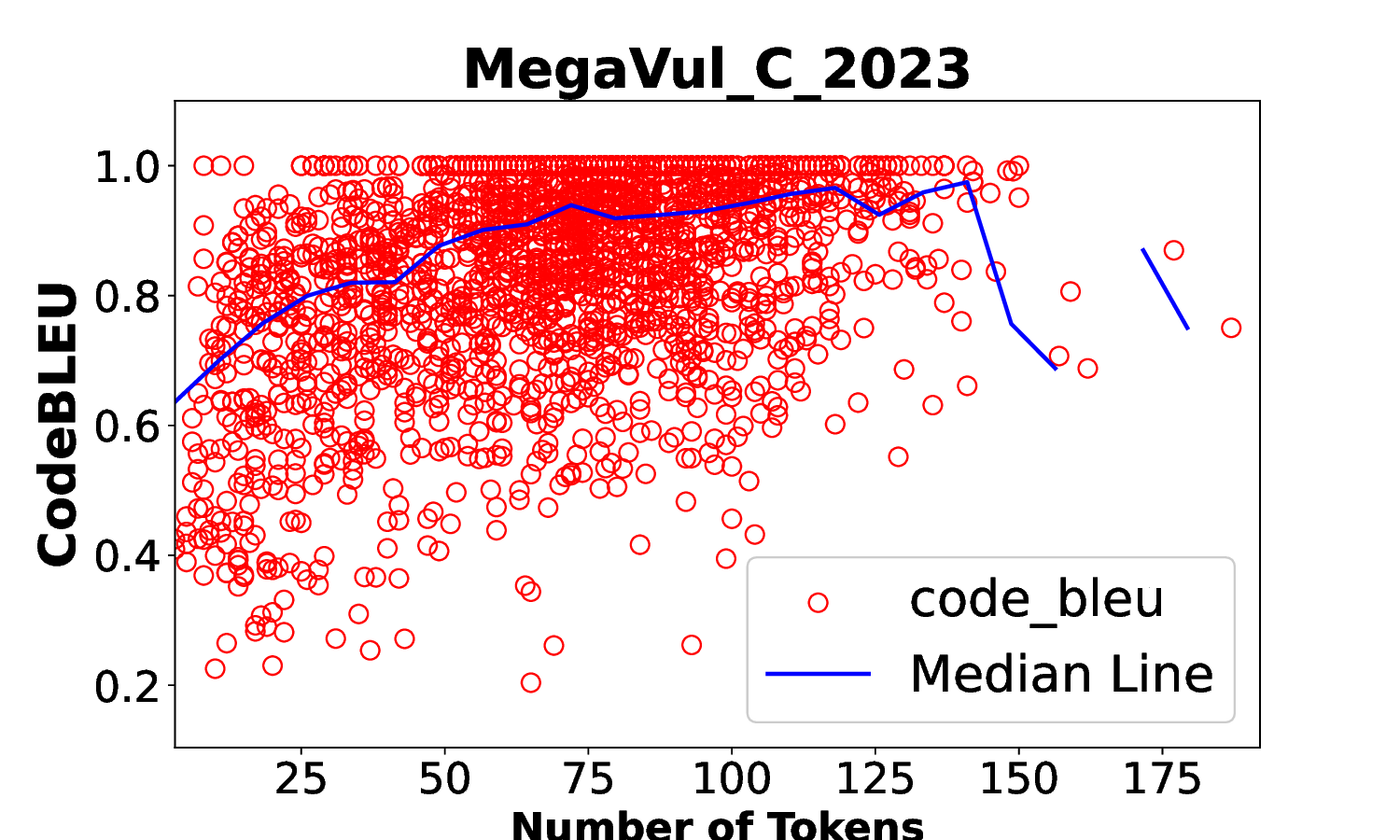}}
      &  \raisebox{-0.5\height}{\includegraphics[width=0.34\textwidth] {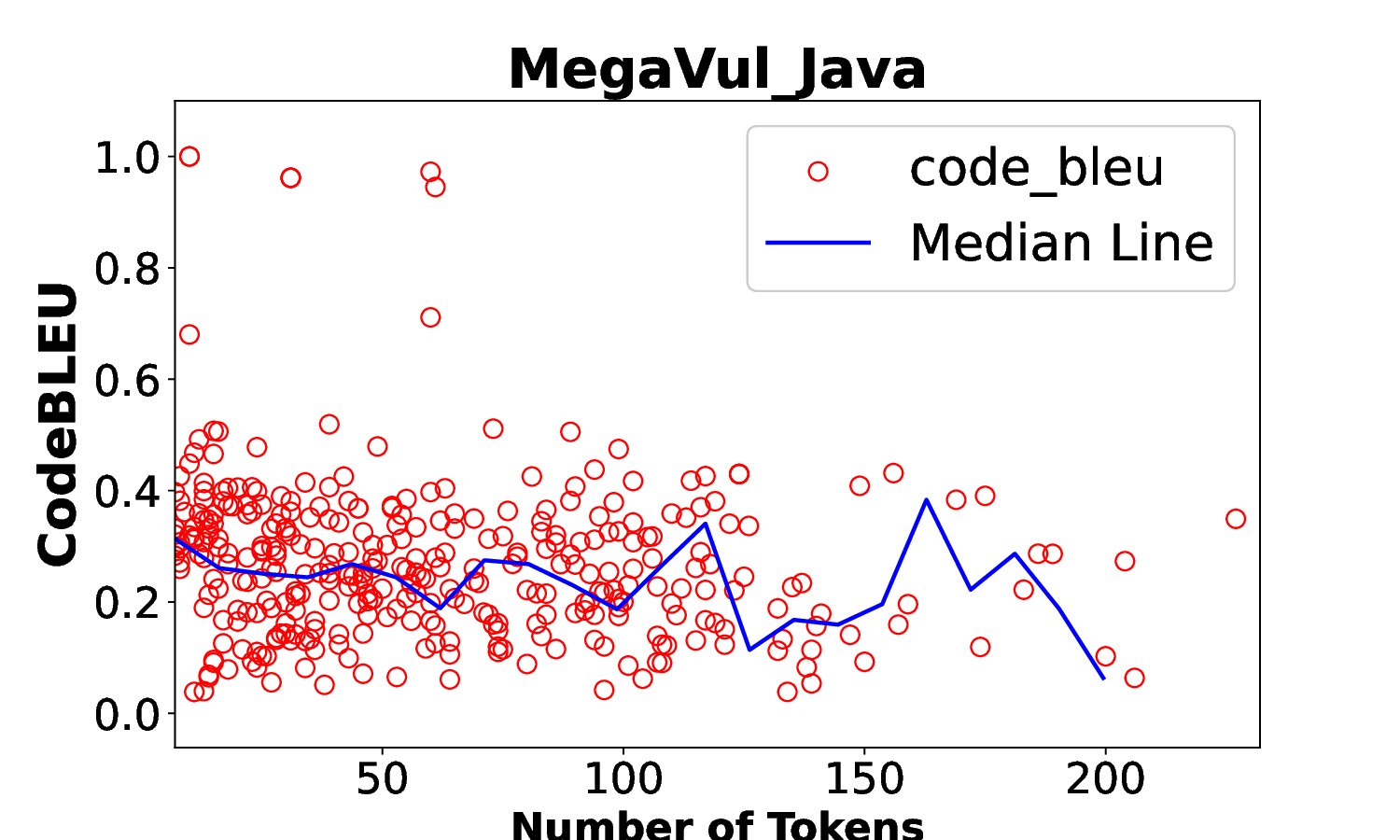}}
      &  \raisebox{-0.5\height}{\includegraphics[width=0.34\textwidth] {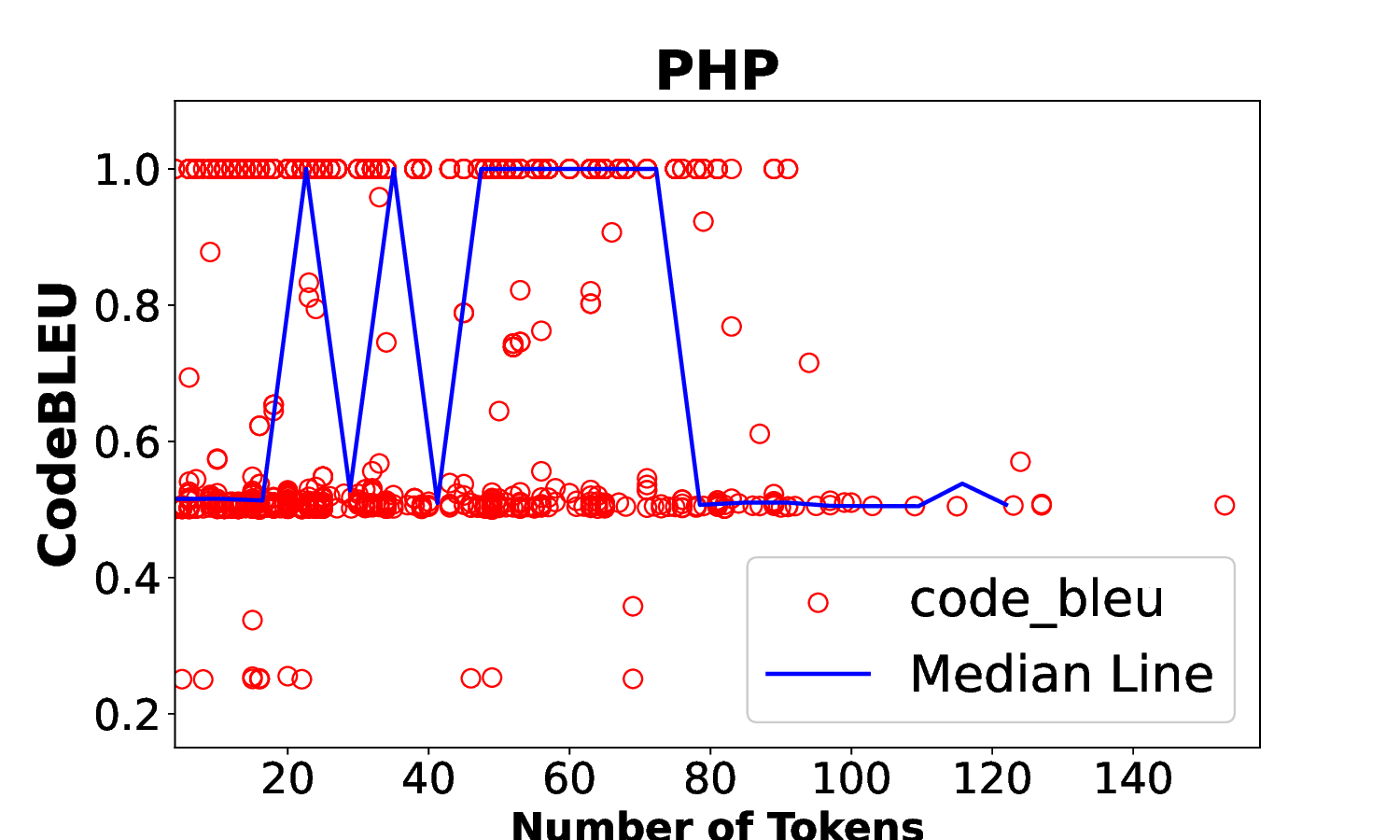}}\\

      &  \raisebox{-0.5\height}{\includegraphics[width=0.34\textwidth] {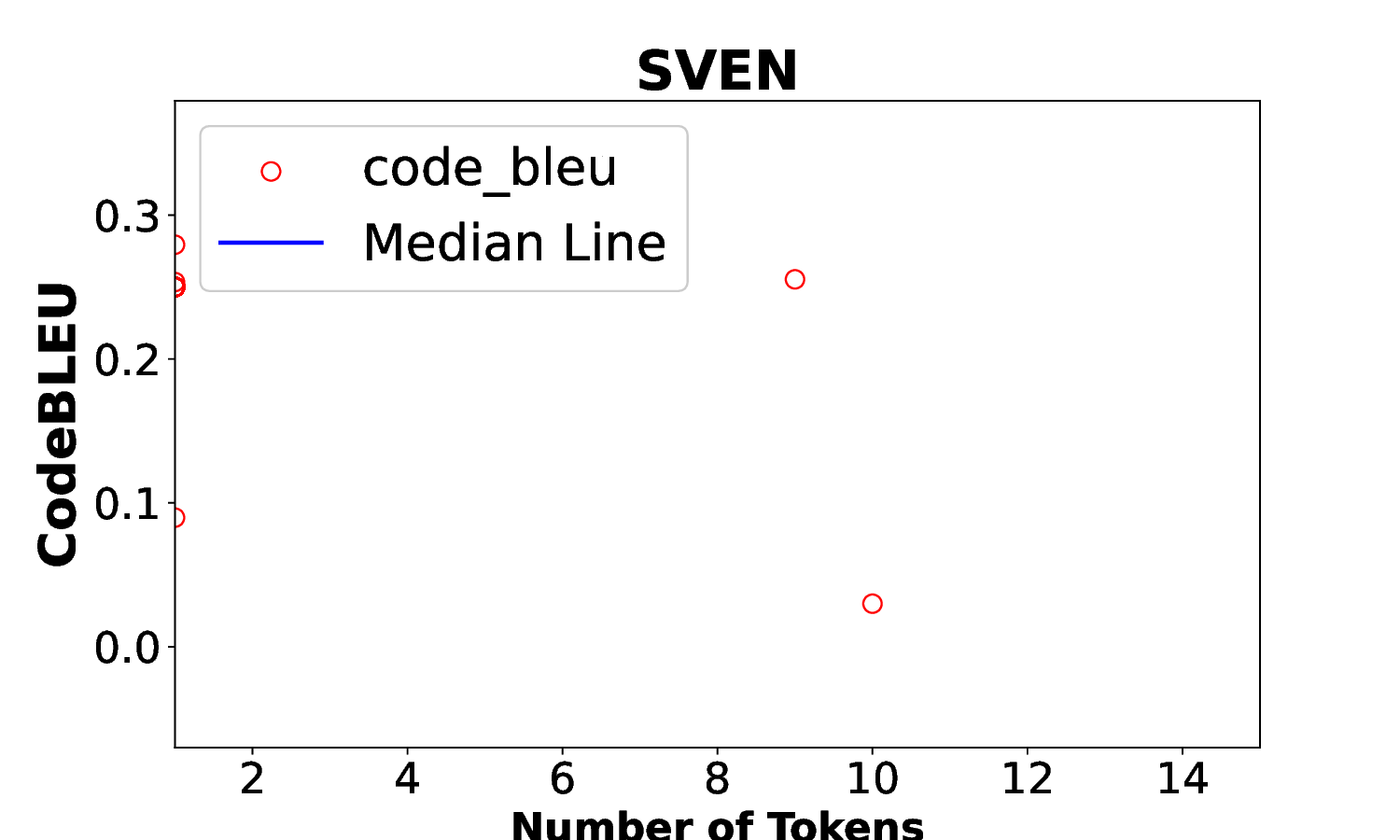}}
      &  \raisebox{-0.5\height}{\includegraphics[width=0.34\textwidth] {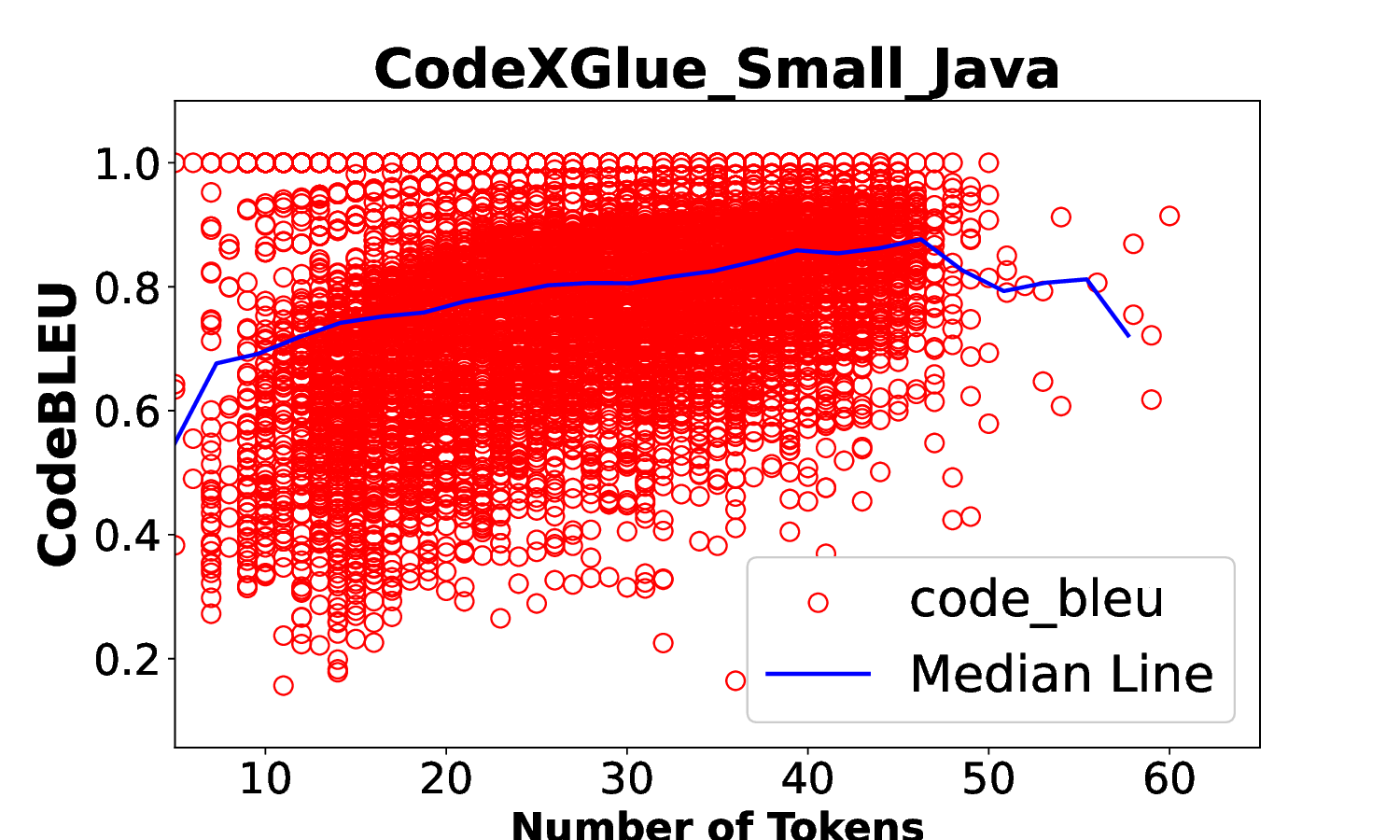}}
      &  \raisebox{-0.5\height}{\includegraphics[width=0.34\textwidth] {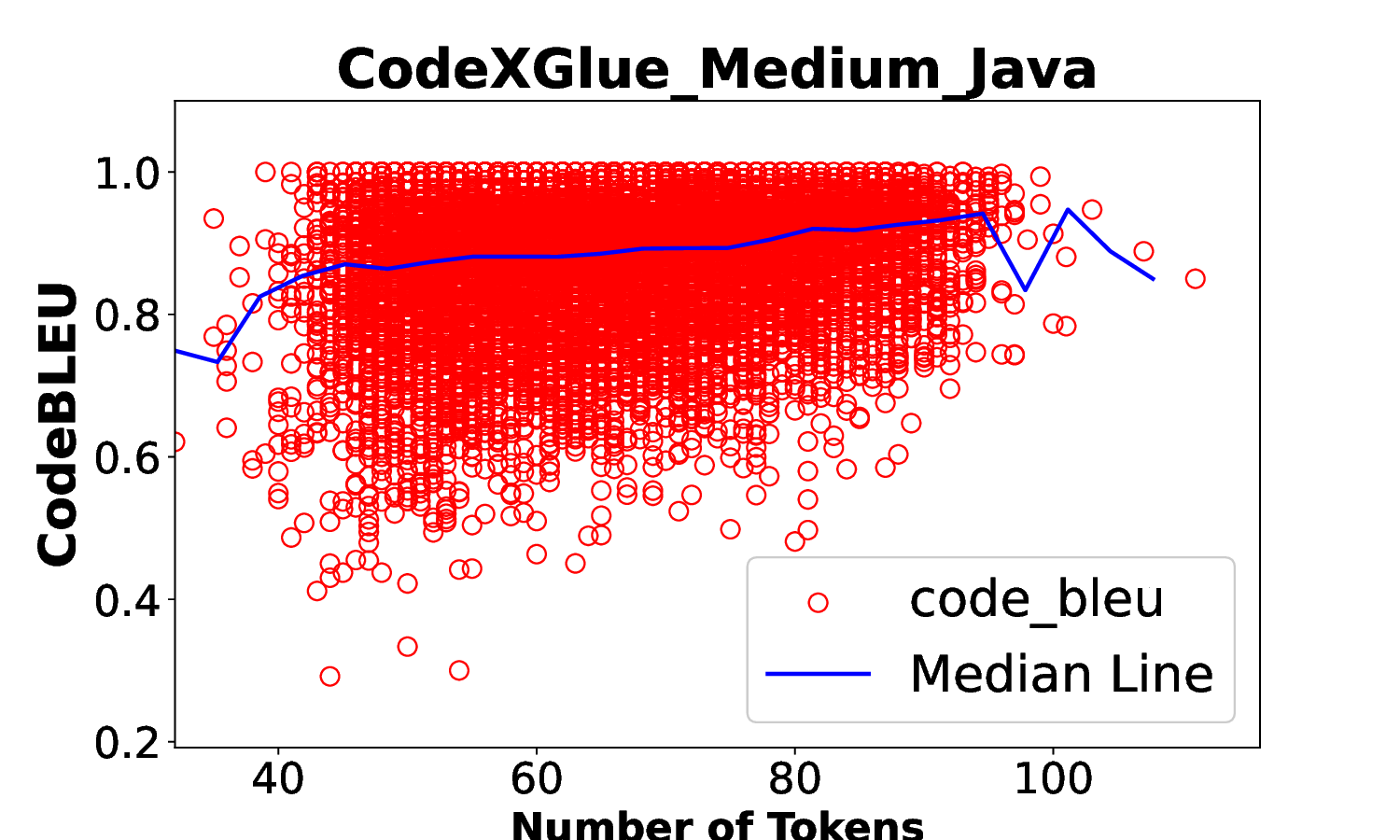}}

    \end{tabularx}
  
  \caption{Correlation Between Patch Length and CodeBLEU (CodeBERT)}
  \label{fig:codebert-codebleu-graph}
\end{figure}

\begin{figure}
  \centering

  \renewcommand{\arraystretch}{1.3}  
    \begin{tabularx}{\textwidth}{
          l X  @{\hspace{3pt}}  X@{\hspace{1pt}}  l 
    }

      & \raisebox{-0.5\height}{\includegraphics[width=0.34\textwidth] {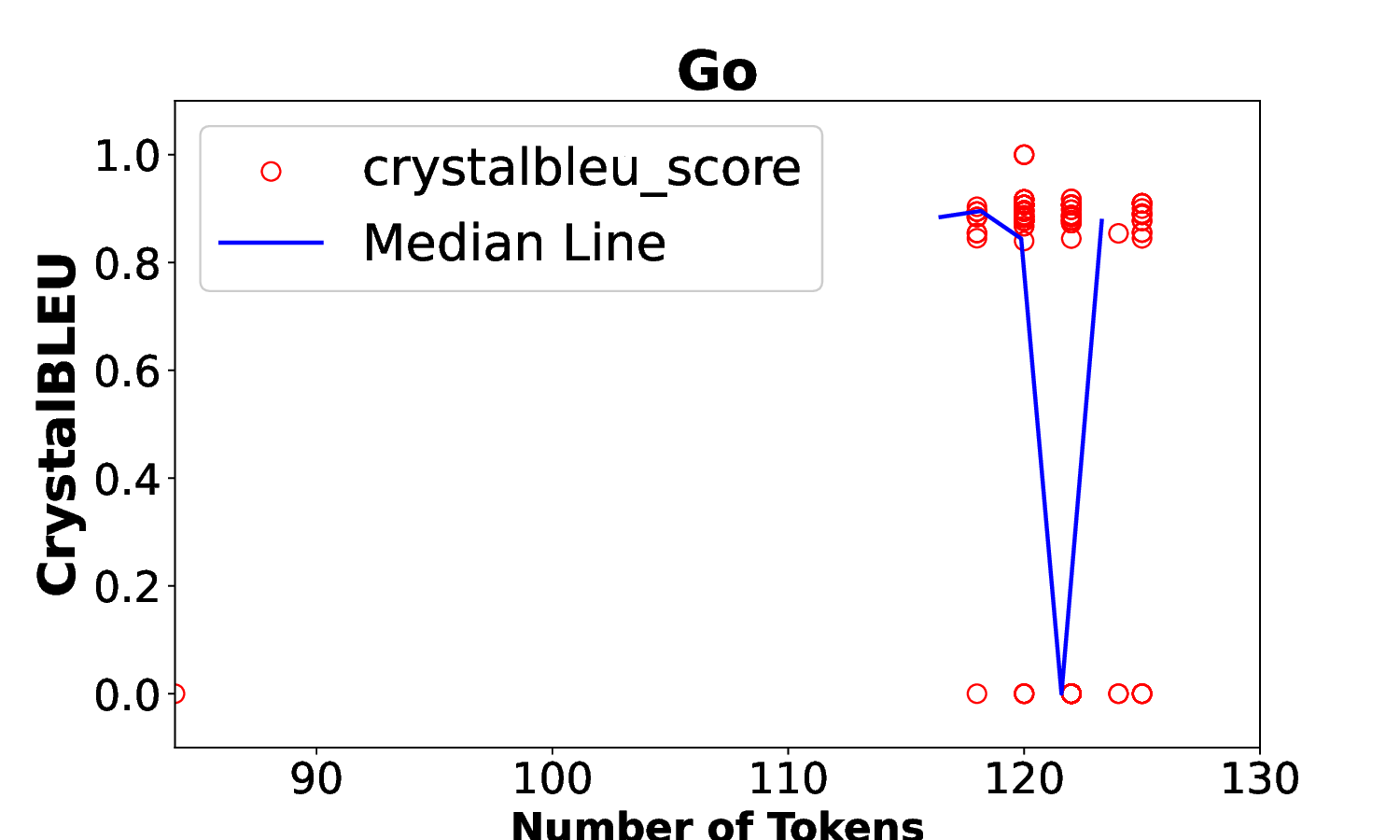}}
      & \raisebox{-0.5\height}{\includegraphics[width=0.34\textwidth] {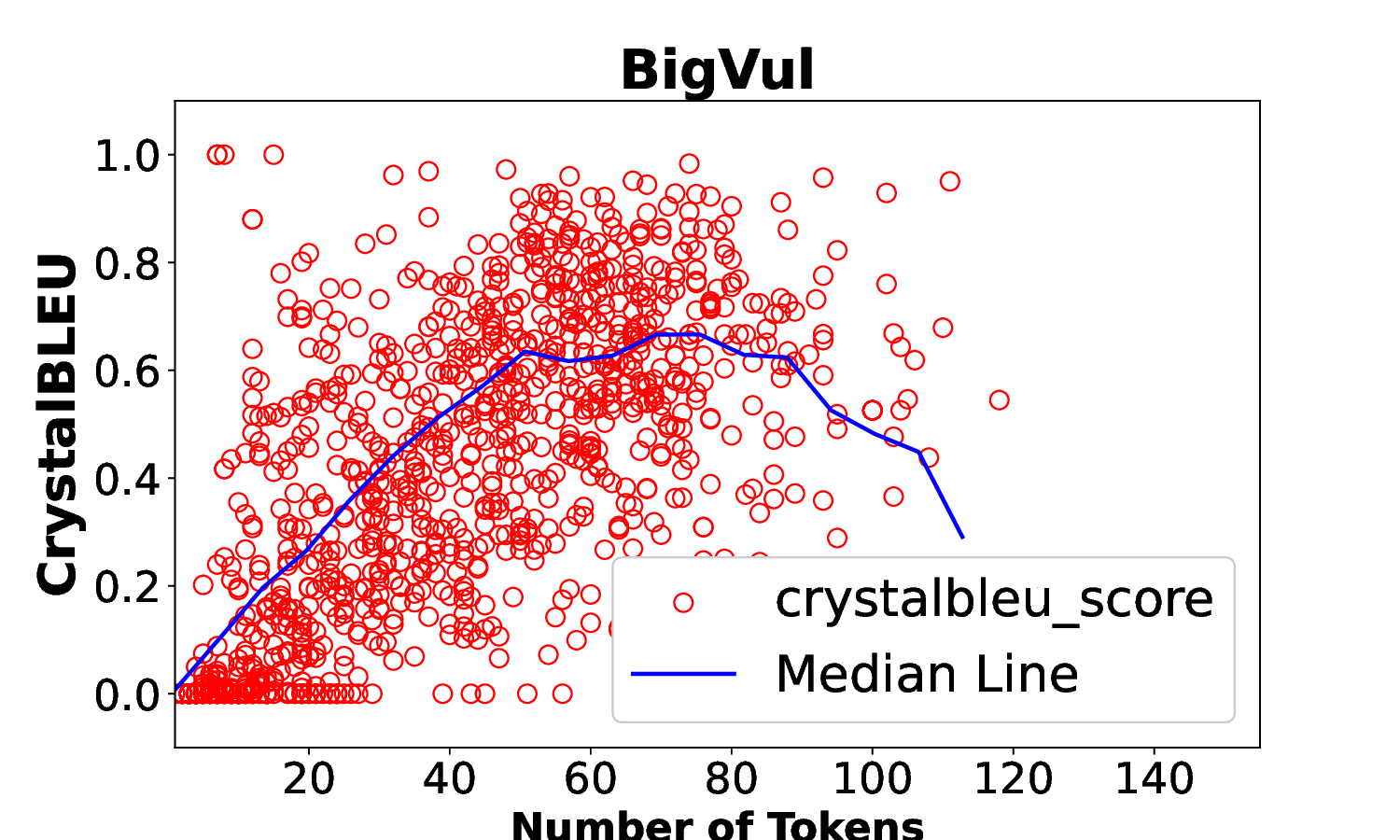}}
       & \raisebox{-0.5\height}{\includegraphics[width=0.34\textwidth] {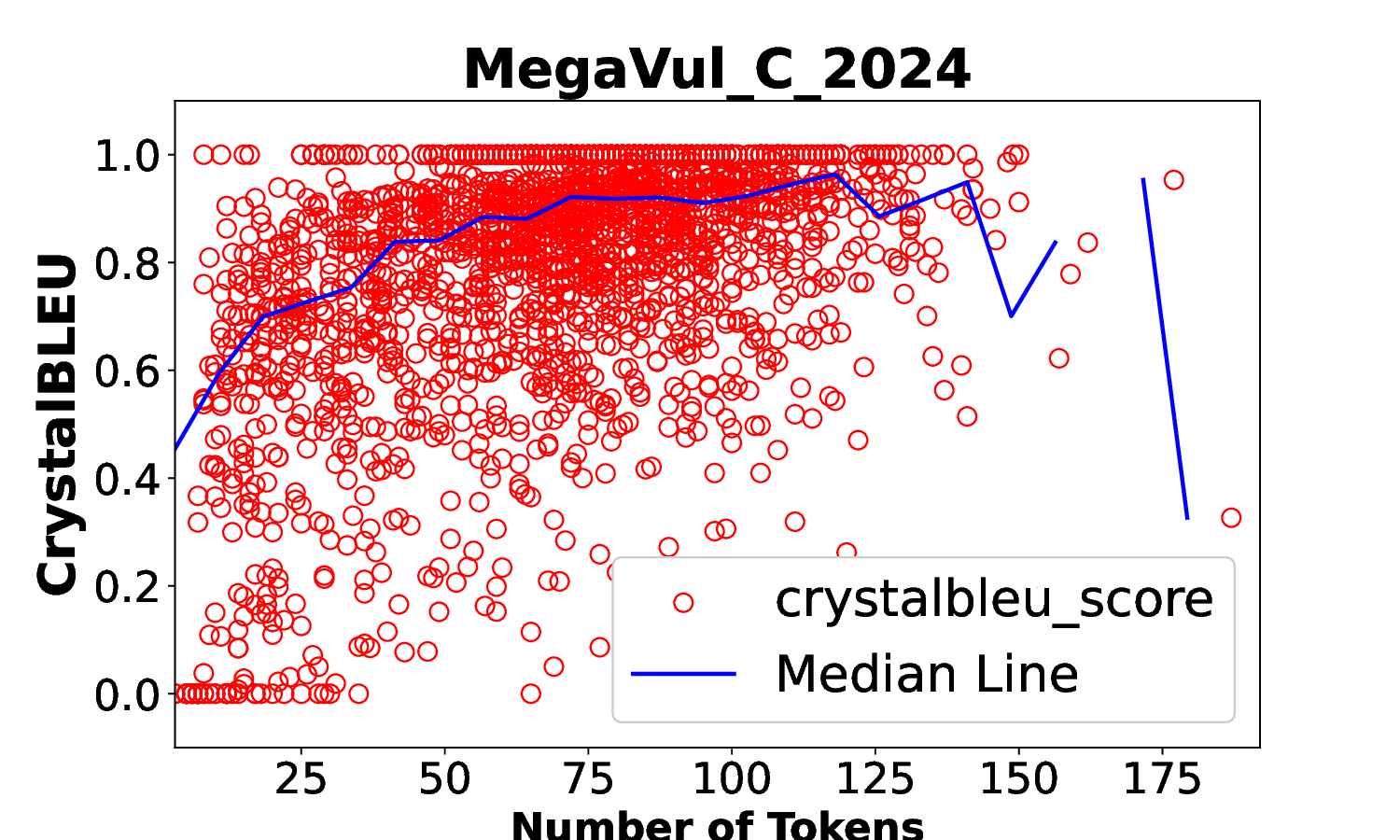}} \\
      
      &  \raisebox{-0.5\height}{\includegraphics[width=0.34\textwidth] {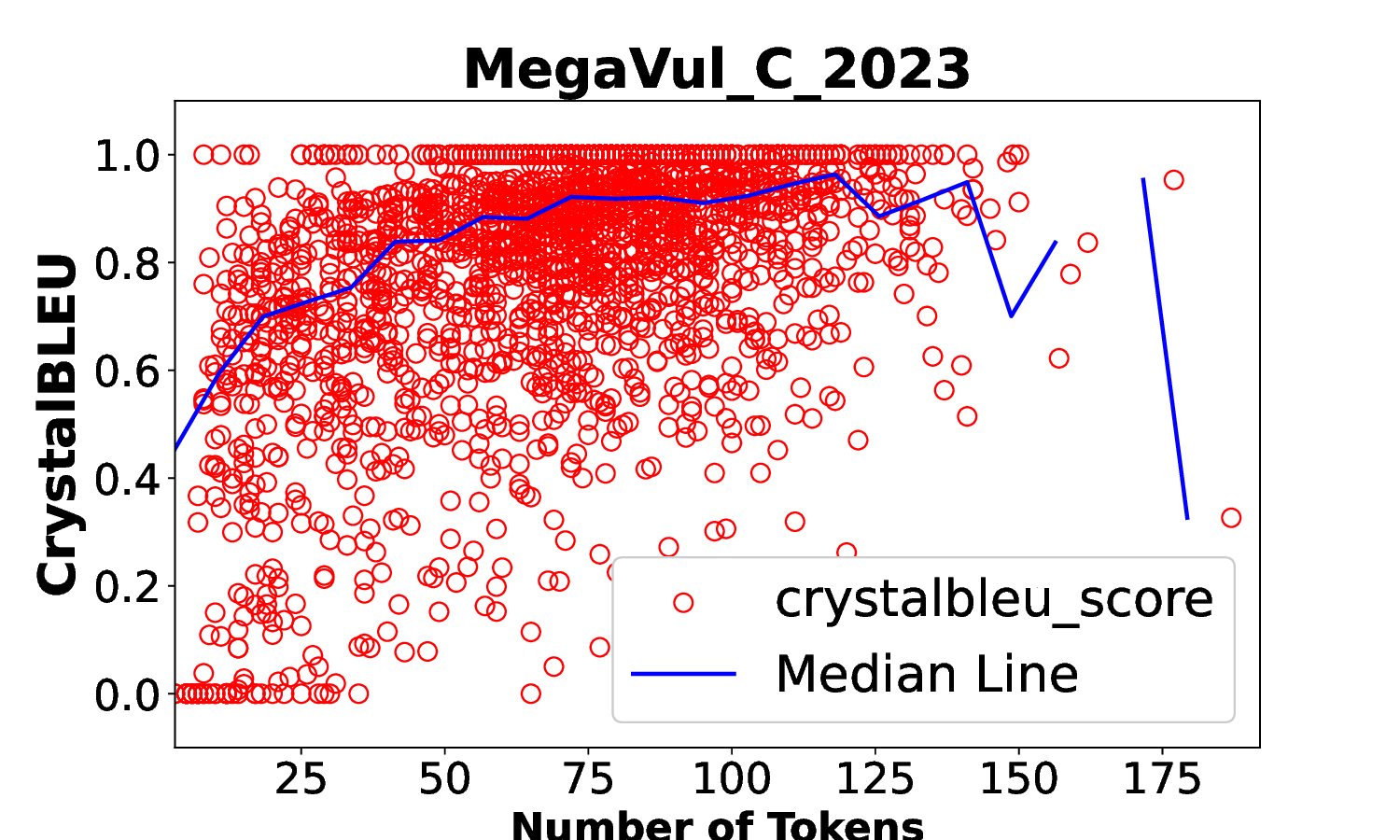}}
      &  \raisebox{-0.5\height}{\includegraphics[width=0.34\textwidth] {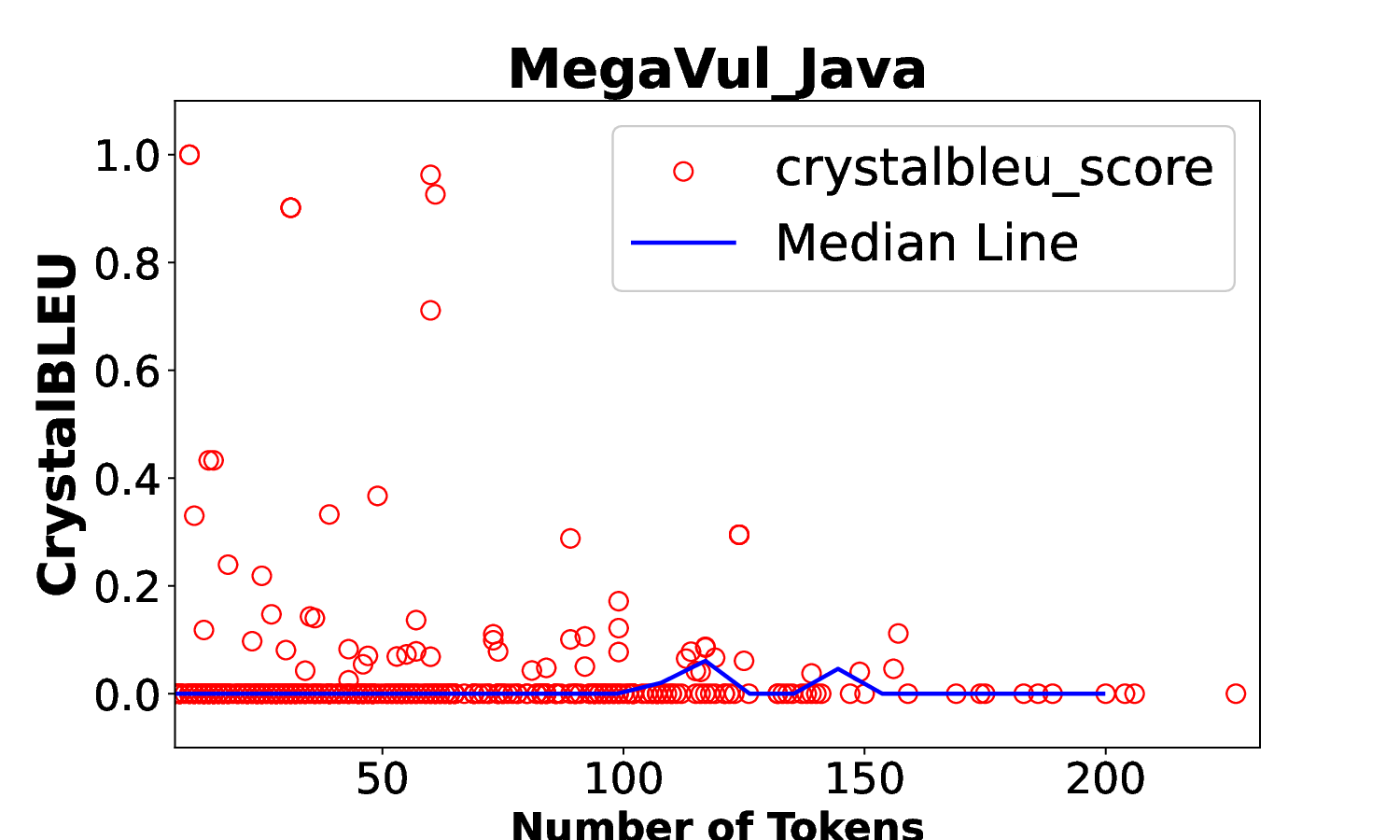}}
      &  \raisebox{-0.5\height}{\includegraphics[width=0.34\textwidth] {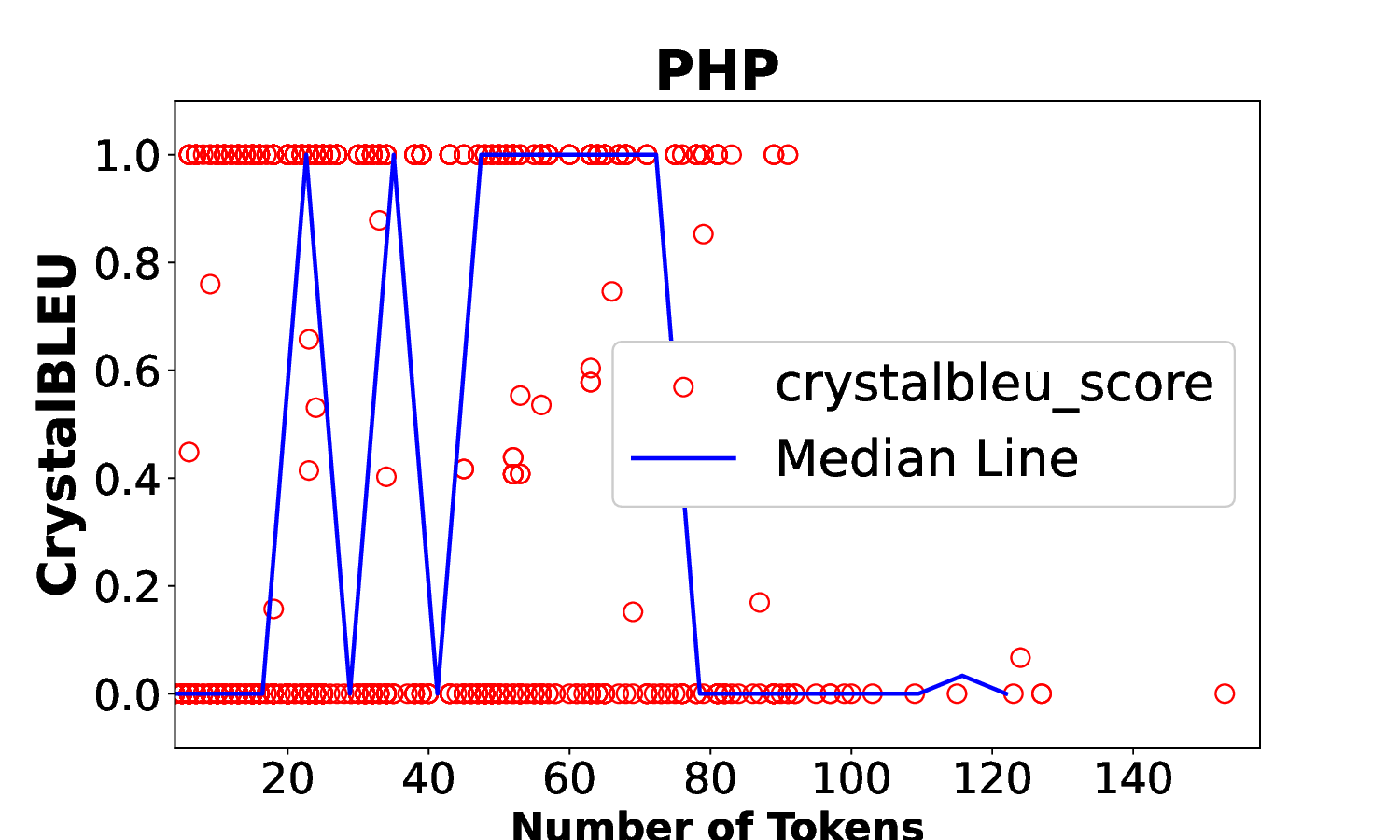}}\\

      &  \raisebox{-0.5\height}{\includegraphics[width=0.34\textwidth] {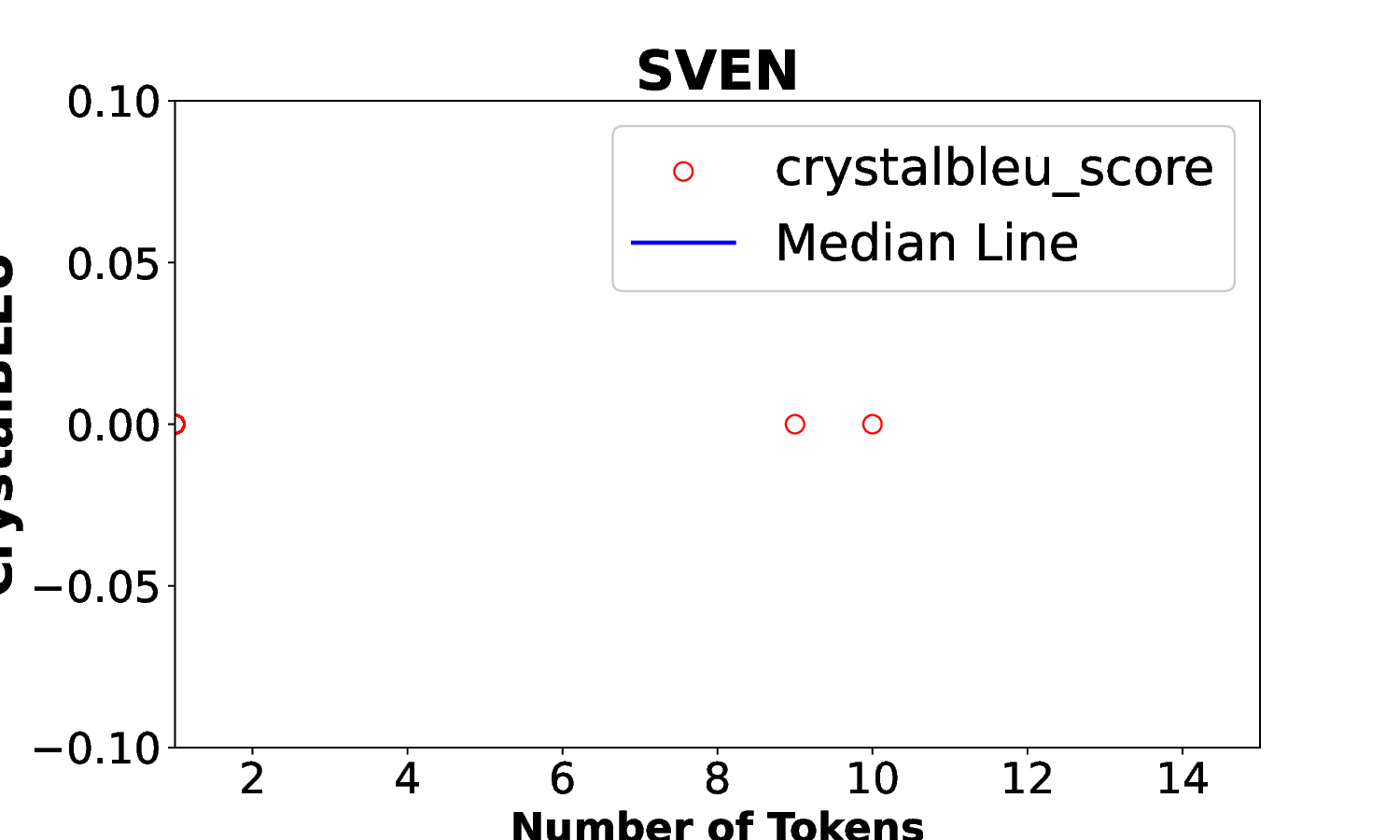}}
      &  \raisebox{-0.5\height}{\includegraphics[width=0.34\textwidth] {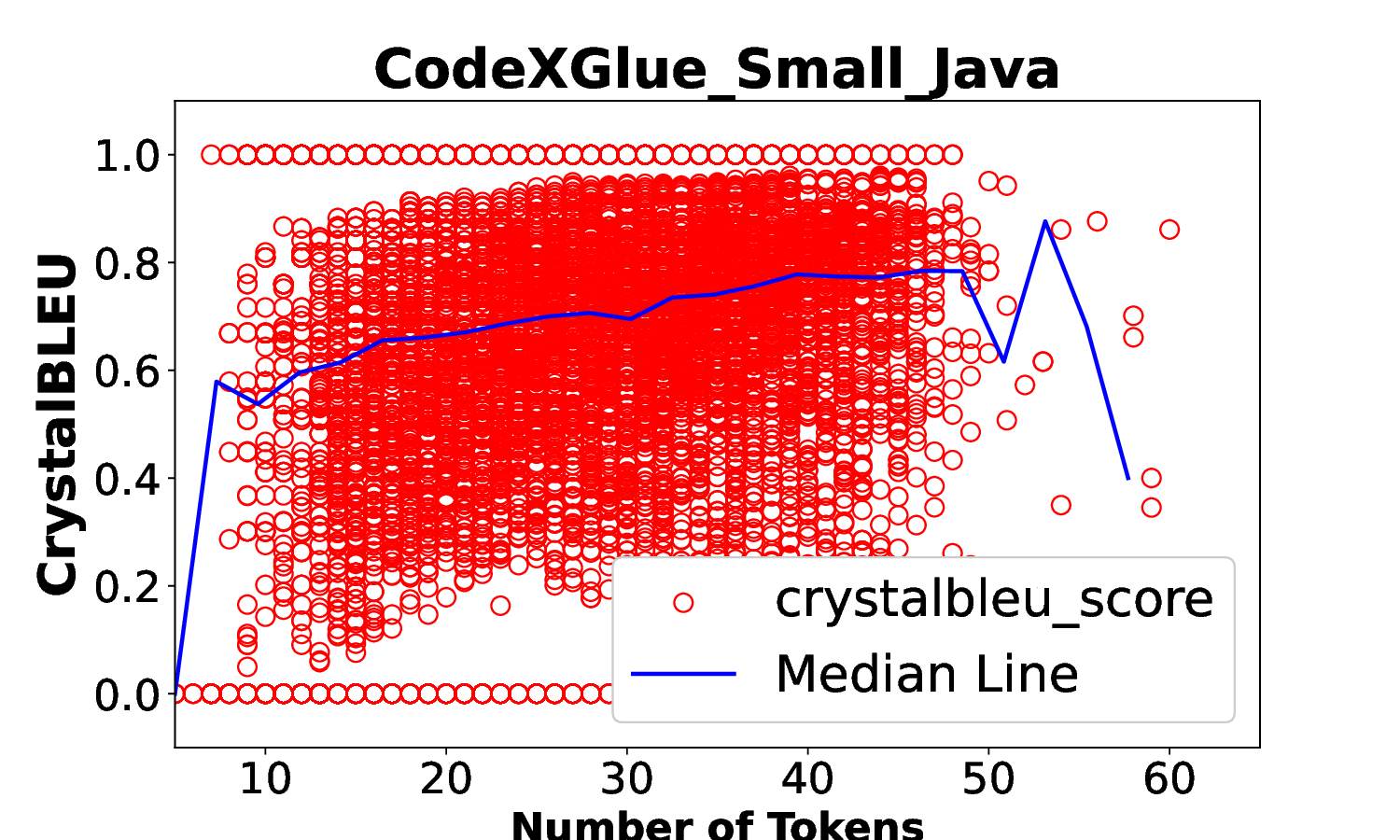}}
      &  \raisebox{-0.5\height}{\includegraphics[width=0.34\textwidth] {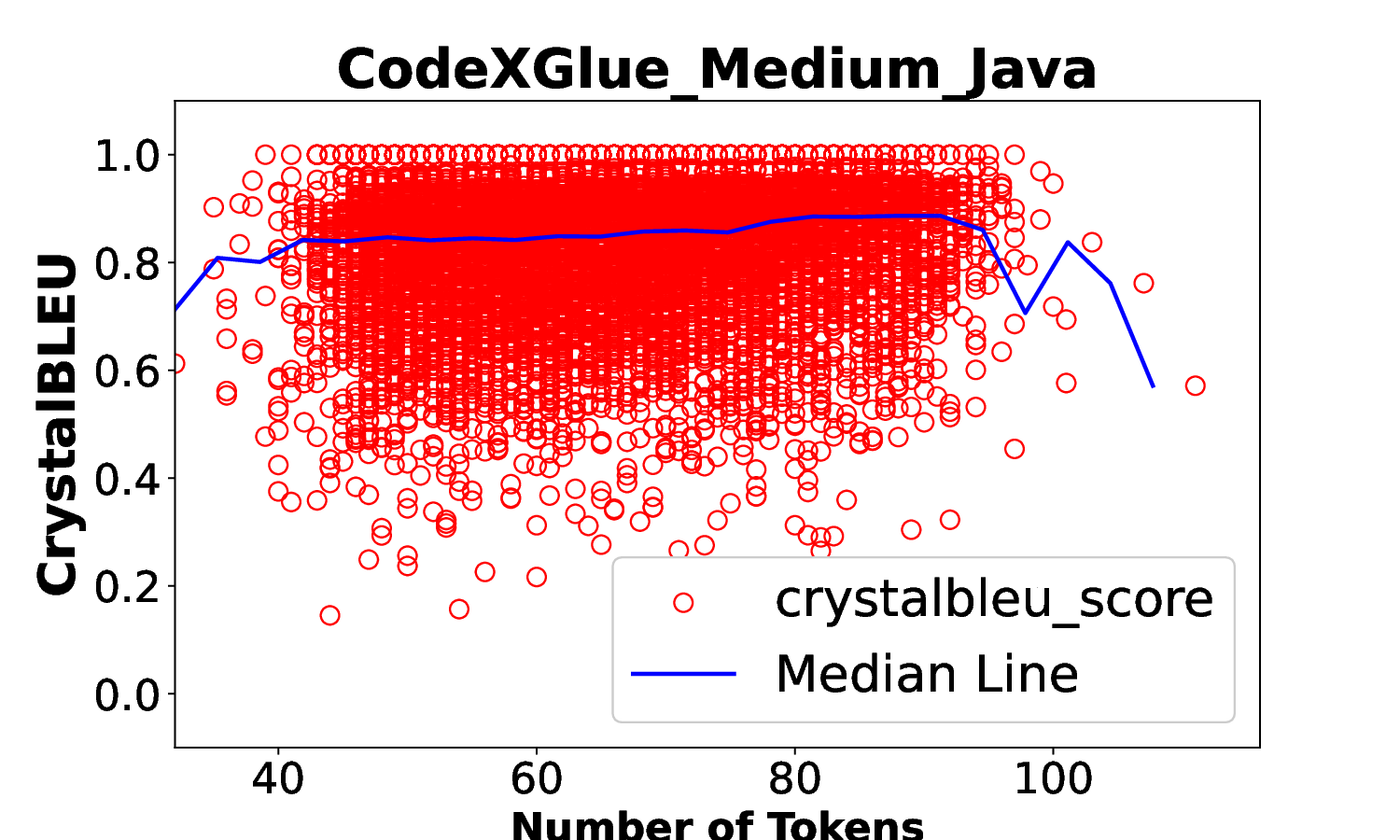}}

    \end{tabularx}
  
  \caption{Correlation Between Patch Length and CrystalBLEU (CodeBERT)}
  \label{fig:codebert-crystalbleu-graph}
\end{figure}

\begin{figure}
  \centering
  
    \begin{tabularx}{\textwidth}{
          l X  @{\hspace{3pt}}  X@{\hspace{1pt}}  l 
    }

      & \raisebox{-0.5\height}{\includegraphics[width=0.34\textwidth] {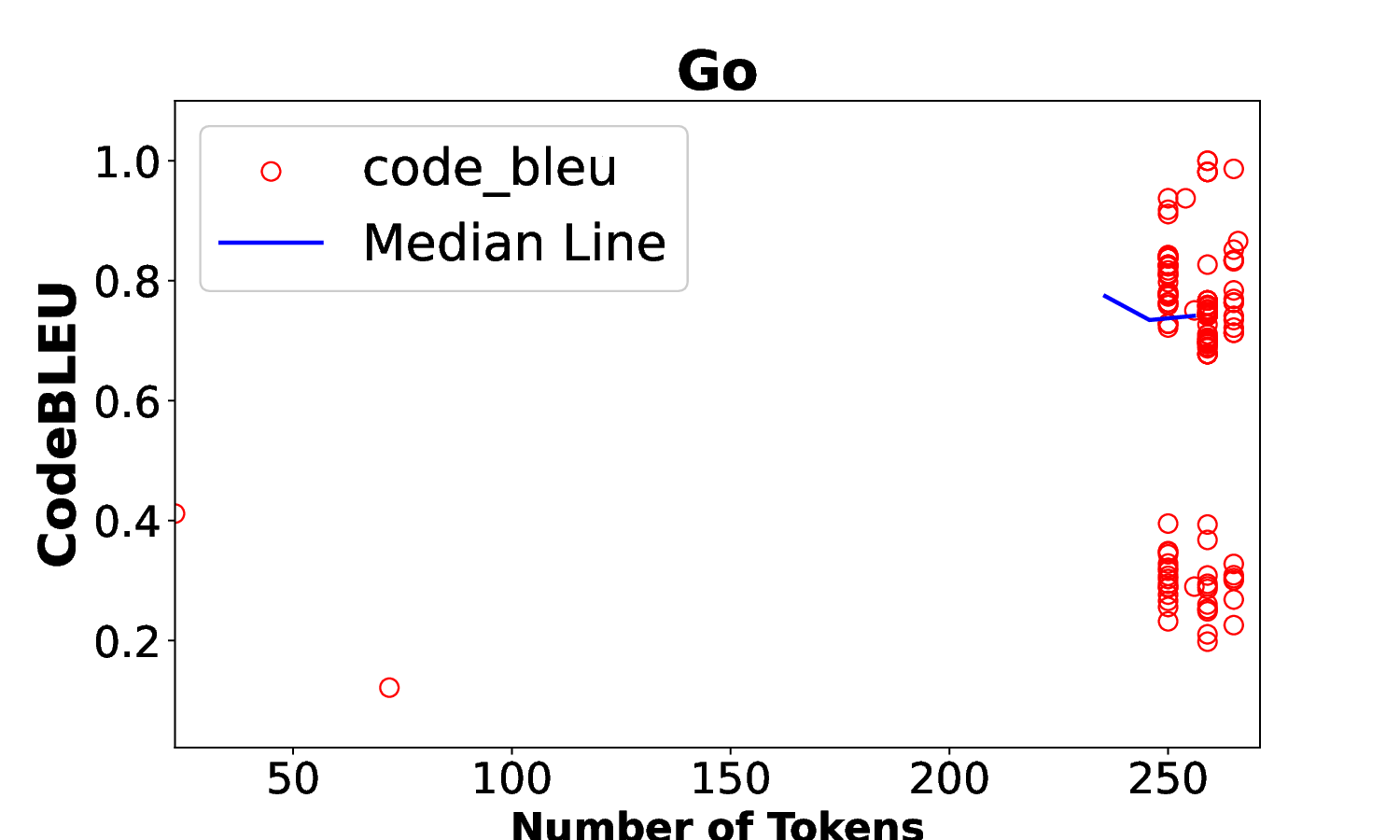}}
      & \raisebox{-0.5\height}{\includegraphics[width=0.34\textwidth] {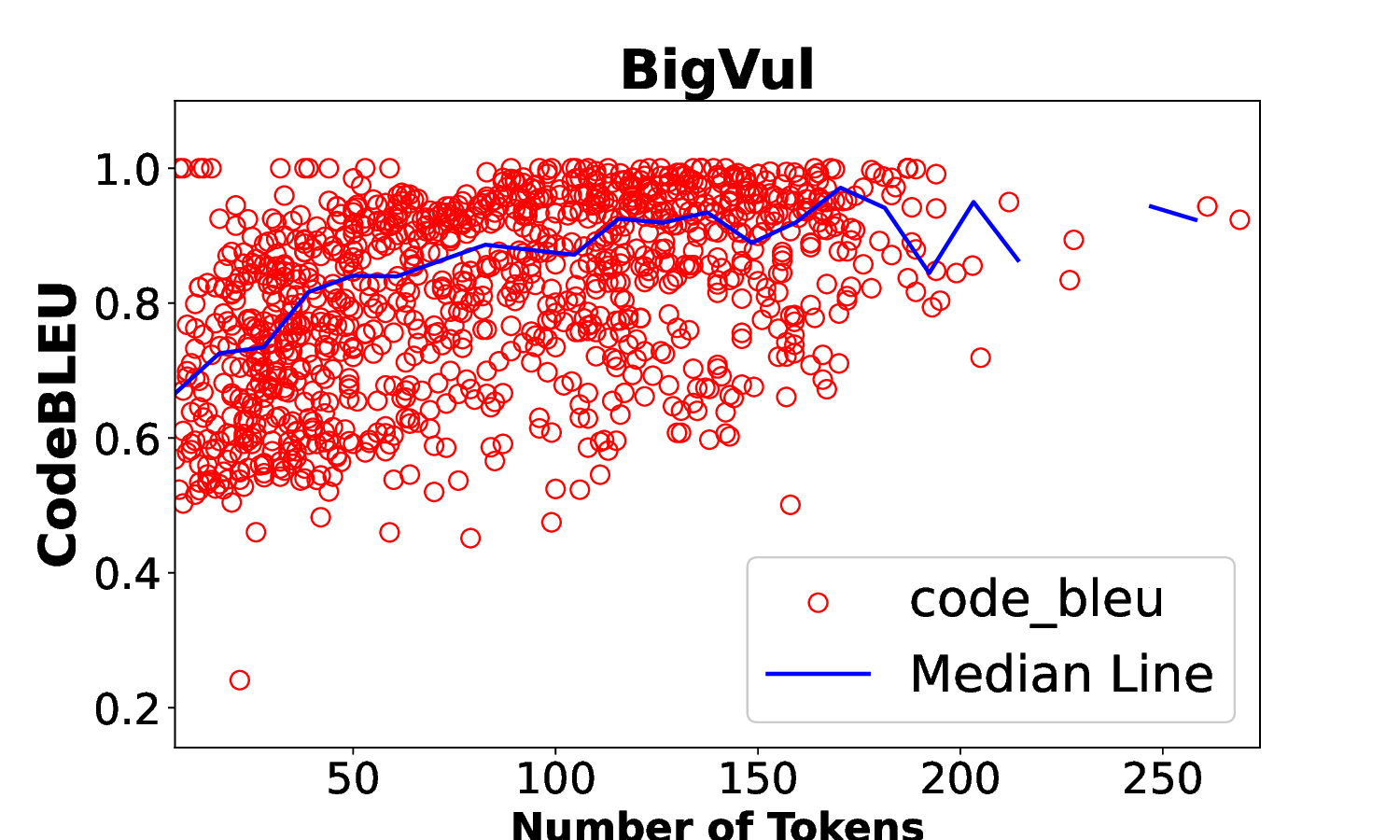}}
       & \raisebox{-0.5\height}{\includegraphics[width=0.34\textwidth] {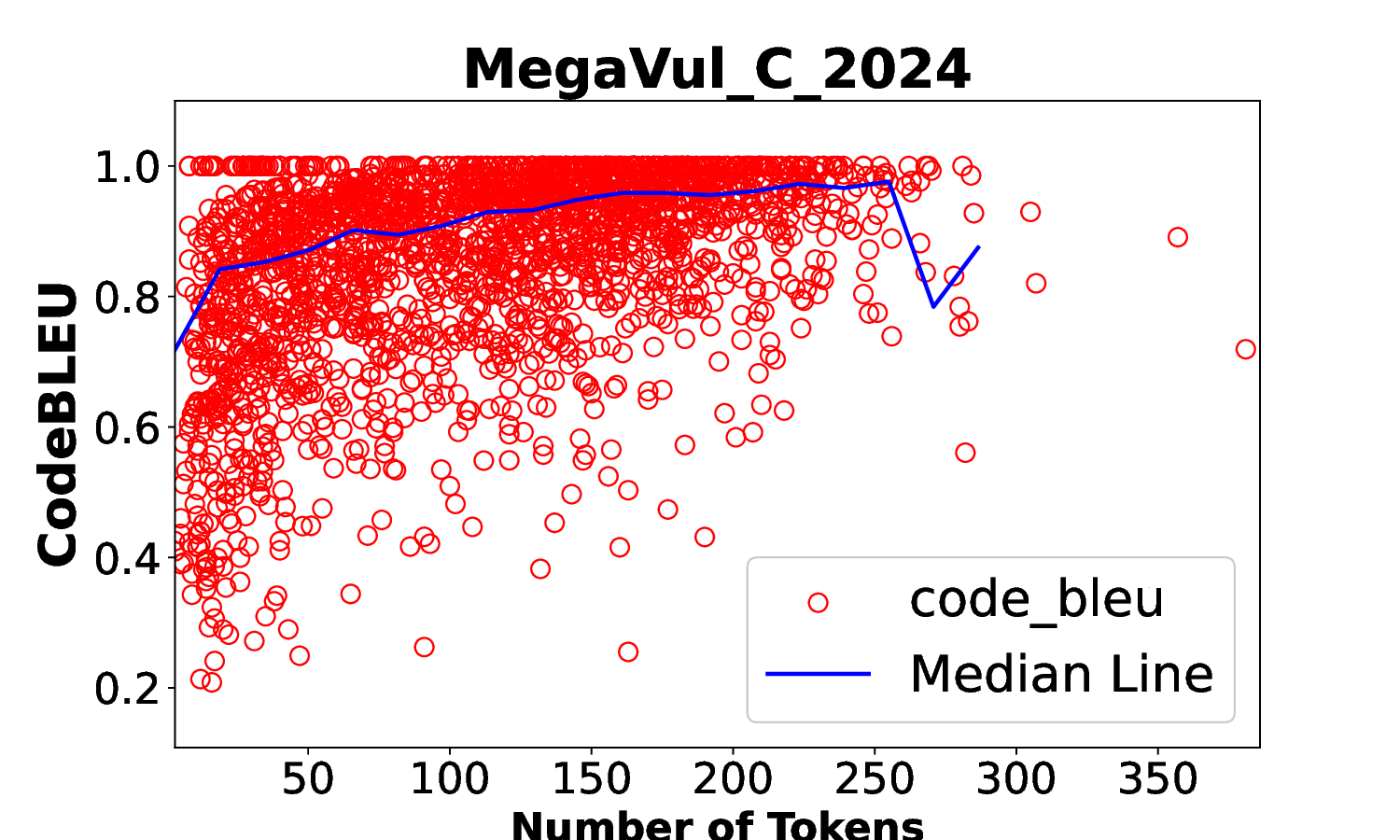}} \\
      
      &  \raisebox{-0.5\height}{\includegraphics[width=0.34\textwidth] {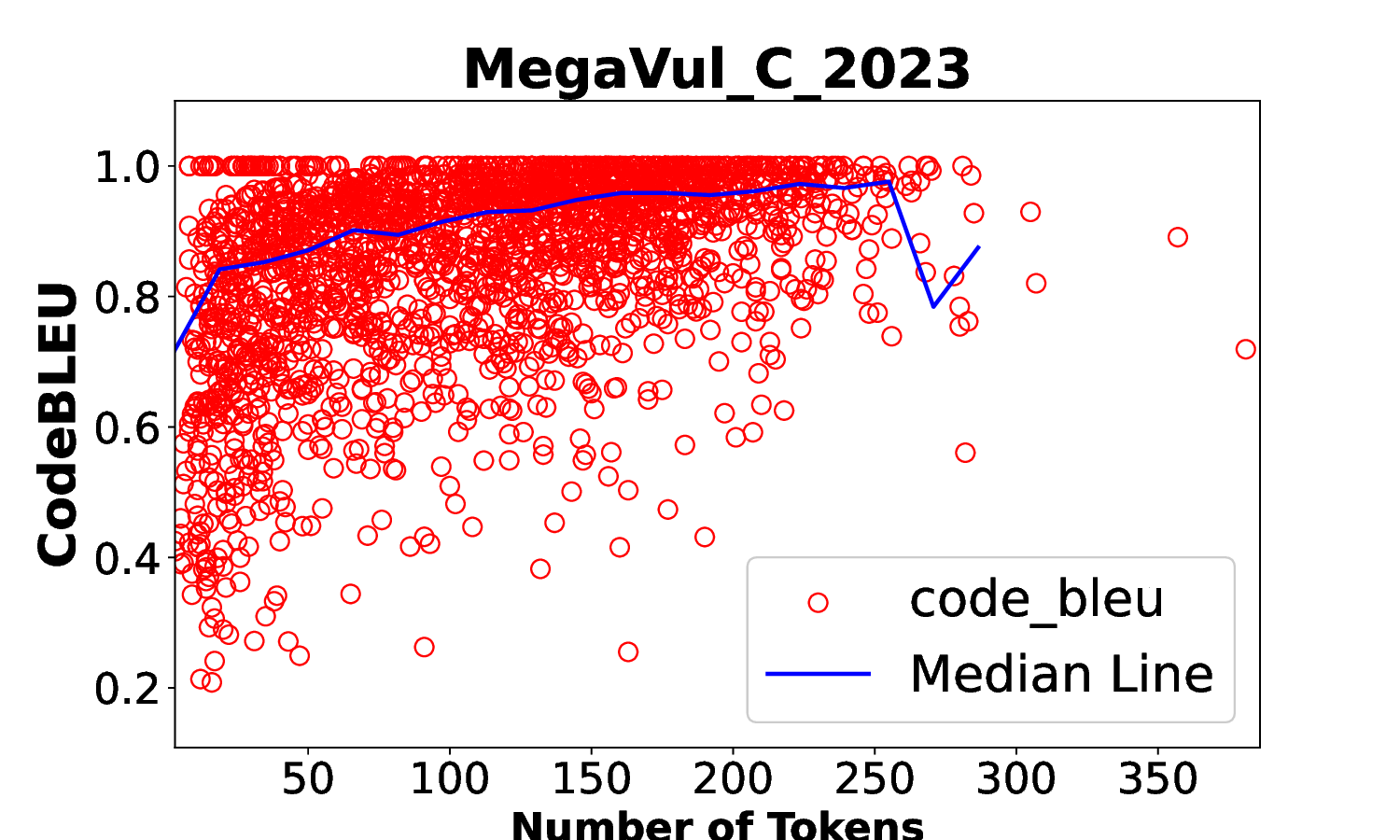}}
      &  \raisebox{-0.5\height}{\includegraphics[width=0.34\textwidth] {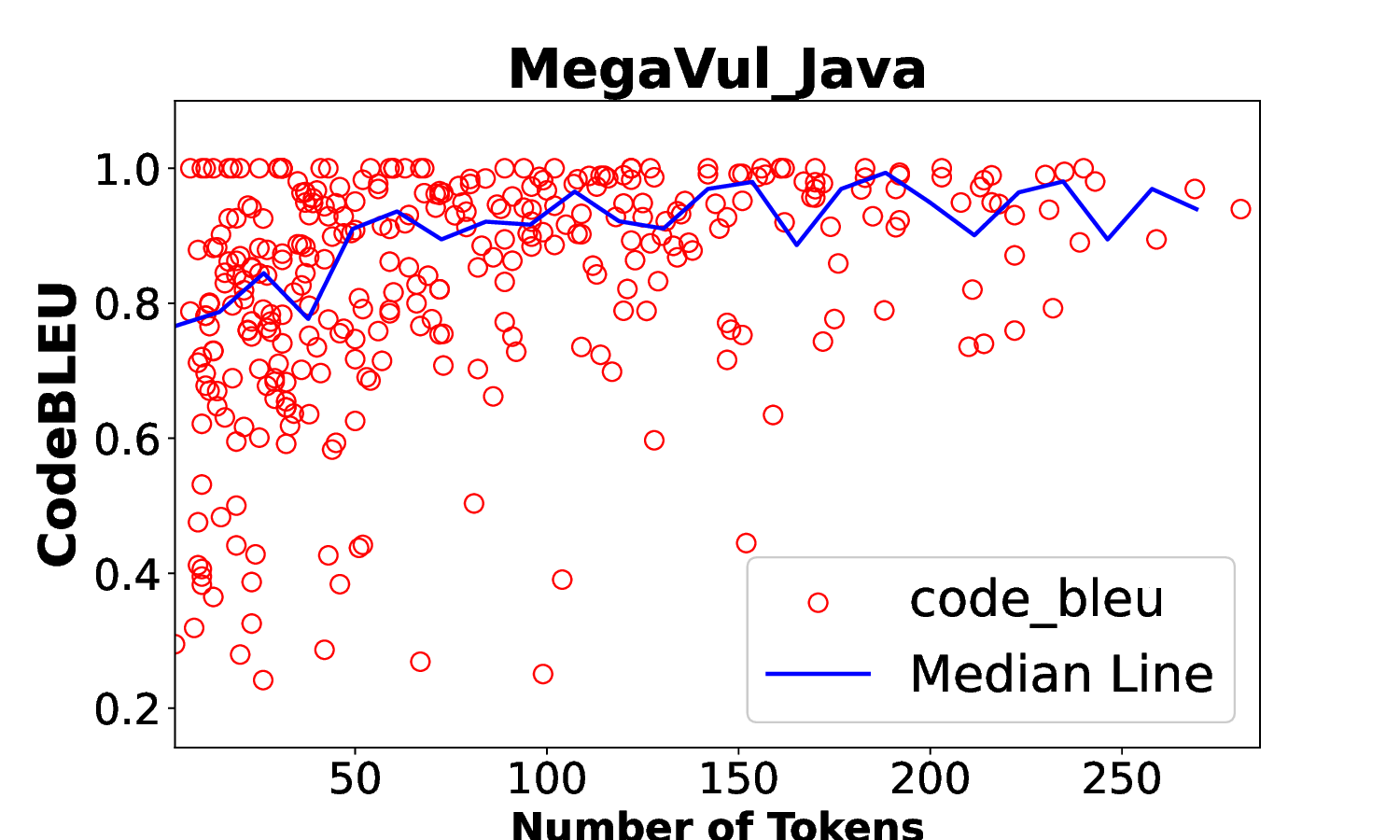}}
      &  \raisebox{-0.5\height}{\includegraphics[width=0.34\textwidth] {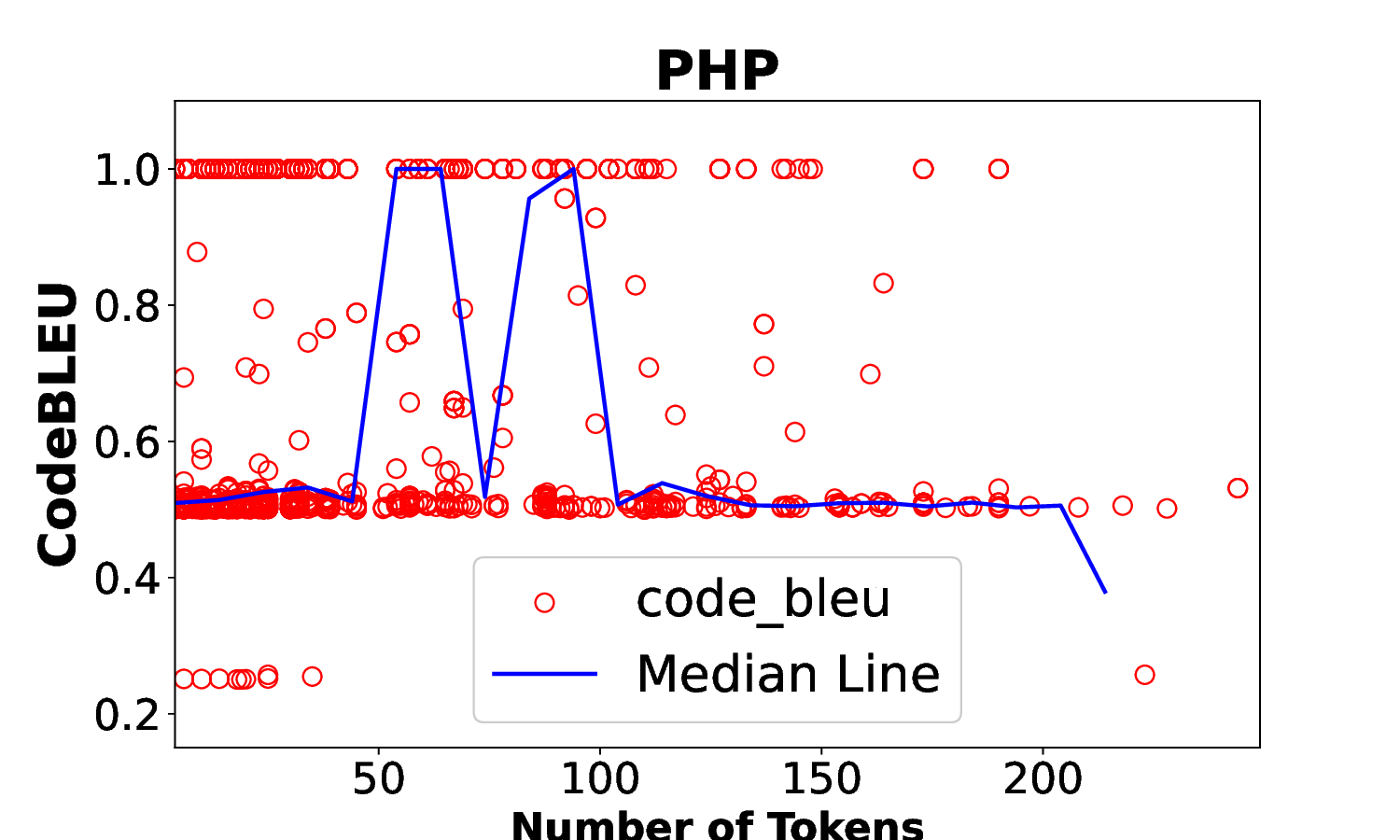}}\\

      &  \raisebox{-0.5\height}{\includegraphics[width=0.34\textwidth] {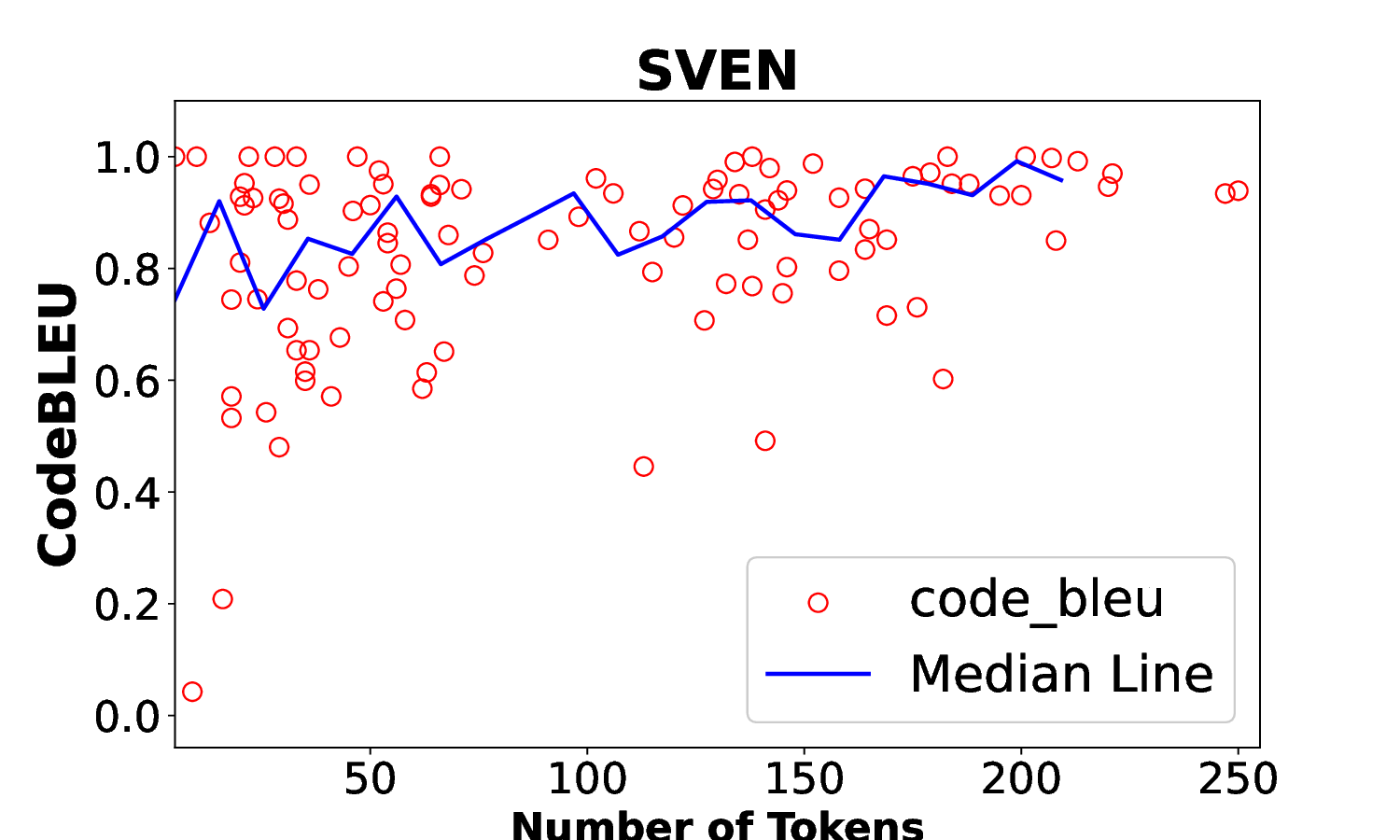}}
      &  \raisebox{-0.5\height}{\includegraphics[width=0.34\textwidth] {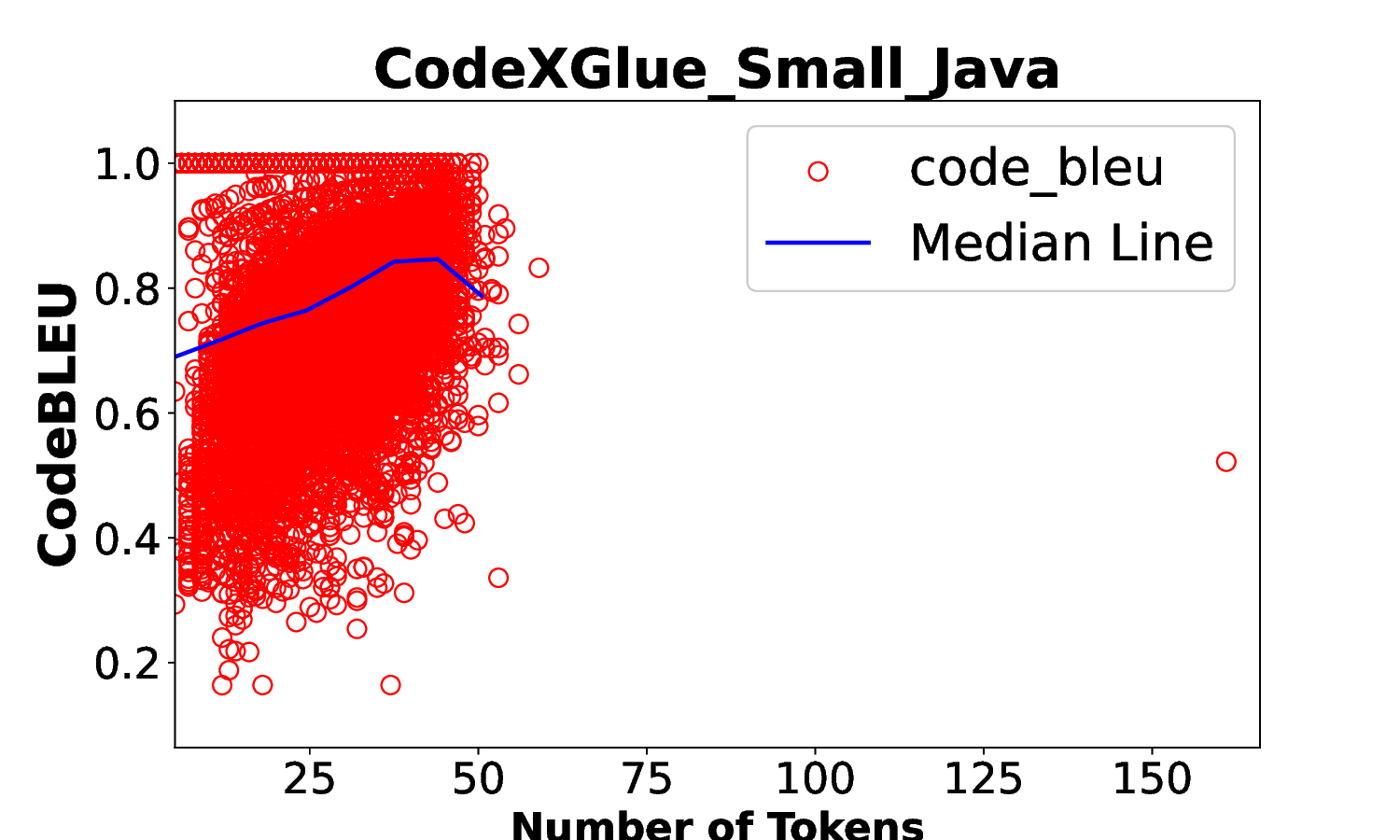}}
      &  \raisebox{-0.5\height}{\includegraphics[width=0.34\textwidth] {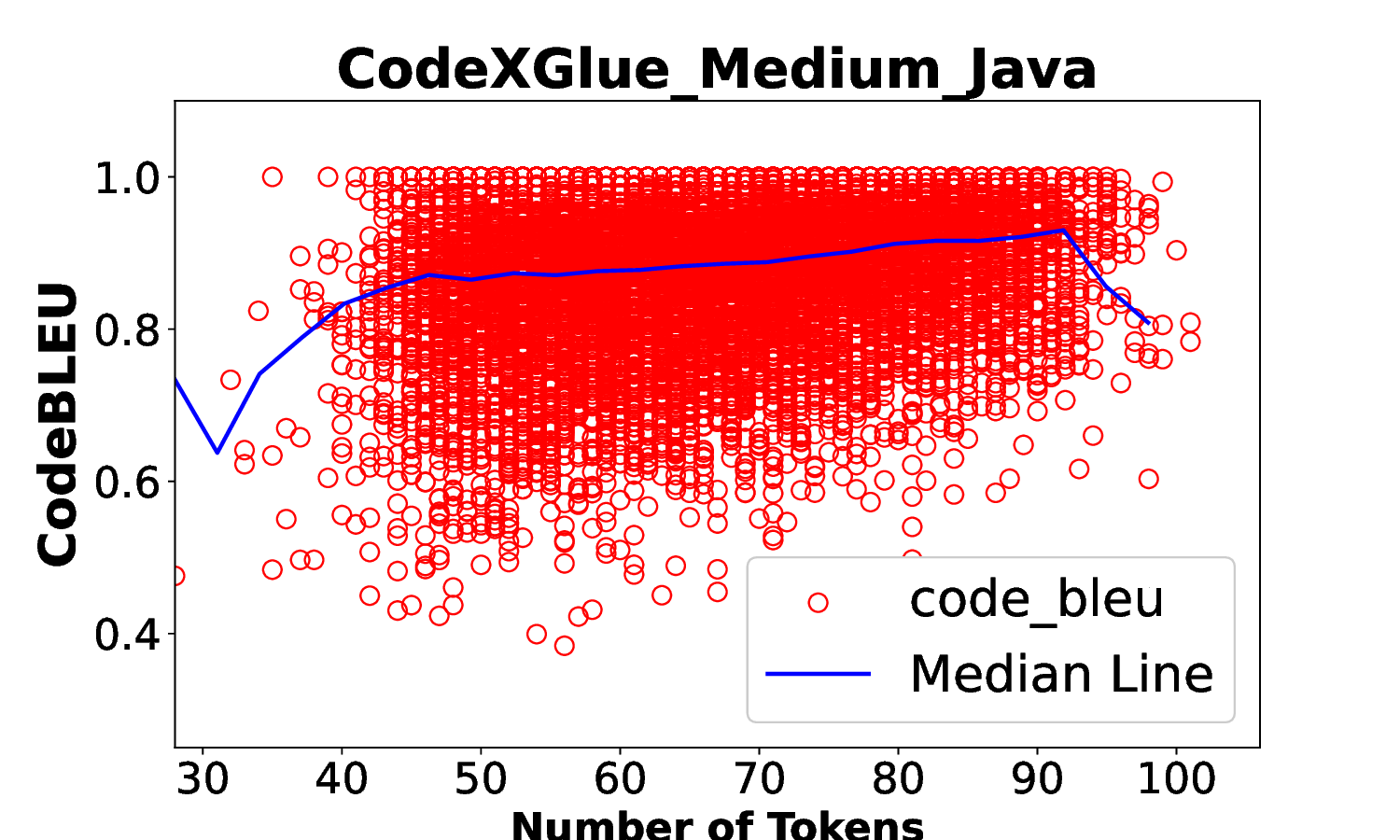}}

    \end{tabularx}
  
  \caption{Correlation Between Patch Length and CodeBLEU (CodeT5)}
  \label{fig:t5-codebleu-graph}
\end{figure}

\begin{figure}
  \centering
  
    \begin{tabularx}{\textwidth}{
          l X  @{\hspace{3pt}}  X@{\hspace{1pt}}  l 
    }

      & \raisebox{-0.5\height}{\includegraphics[width=0.34\textwidth] {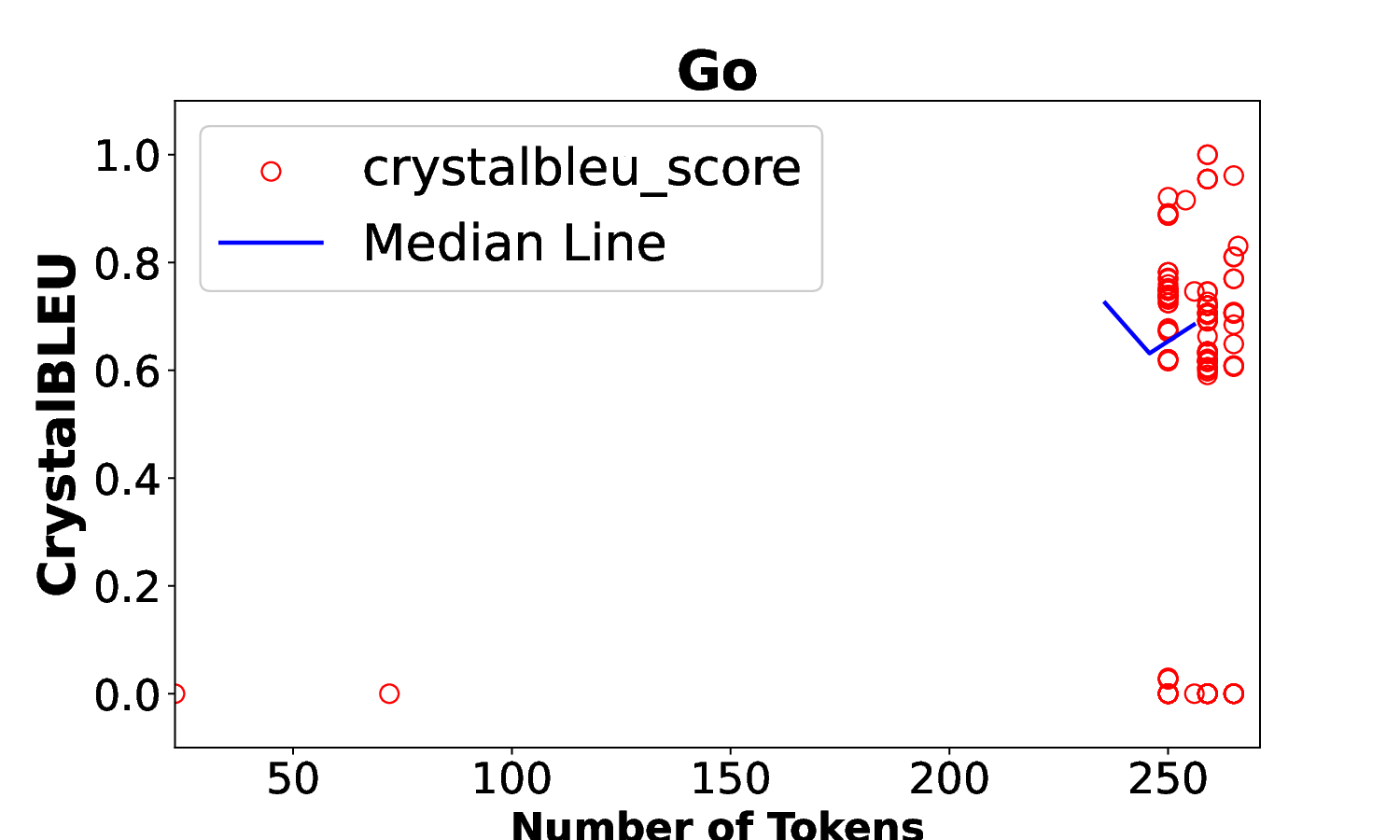}}
      & \raisebox{-0.5\height}{\includegraphics[width=0.34\textwidth] {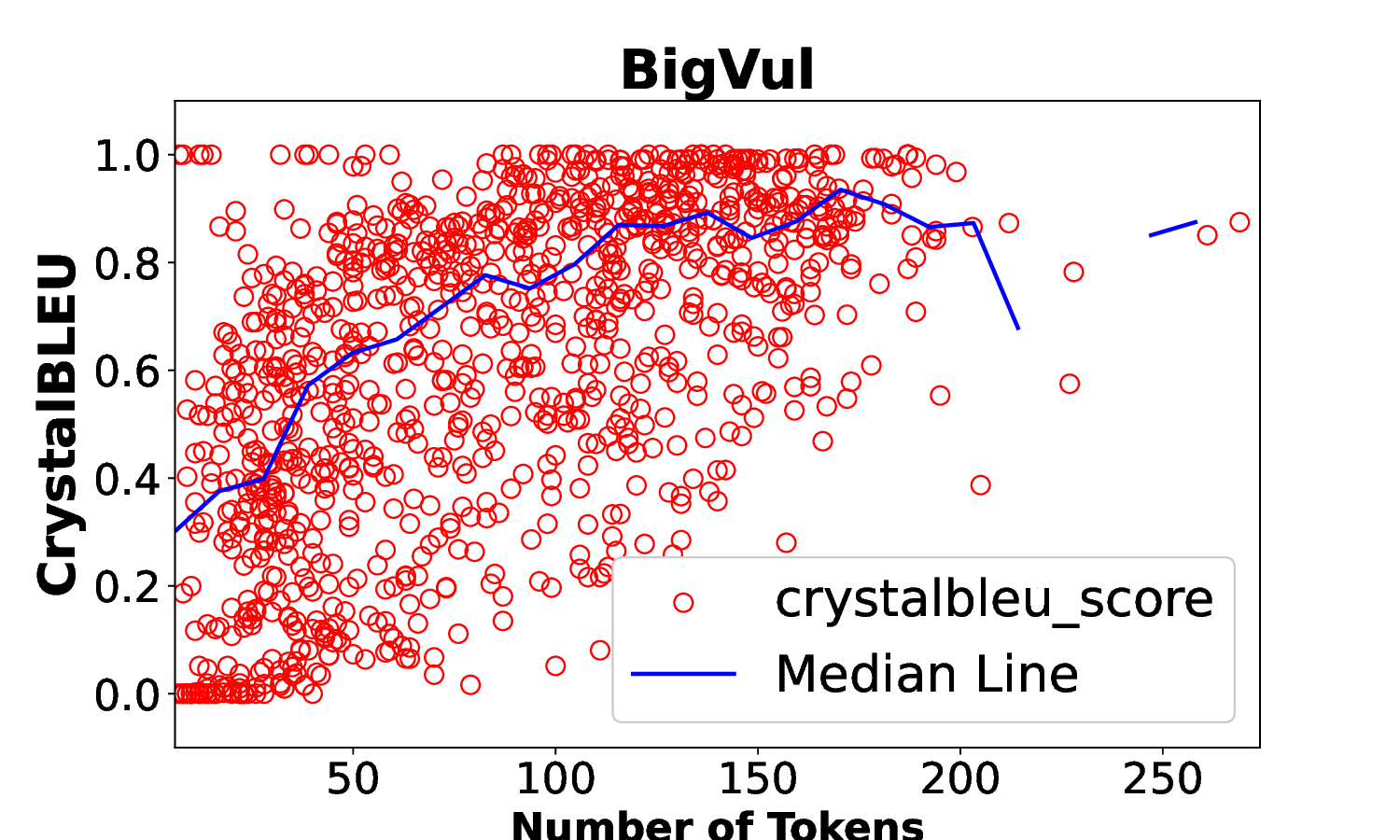}}
       & \raisebox{-0.5\height}{\includegraphics[width=0.34\textwidth] {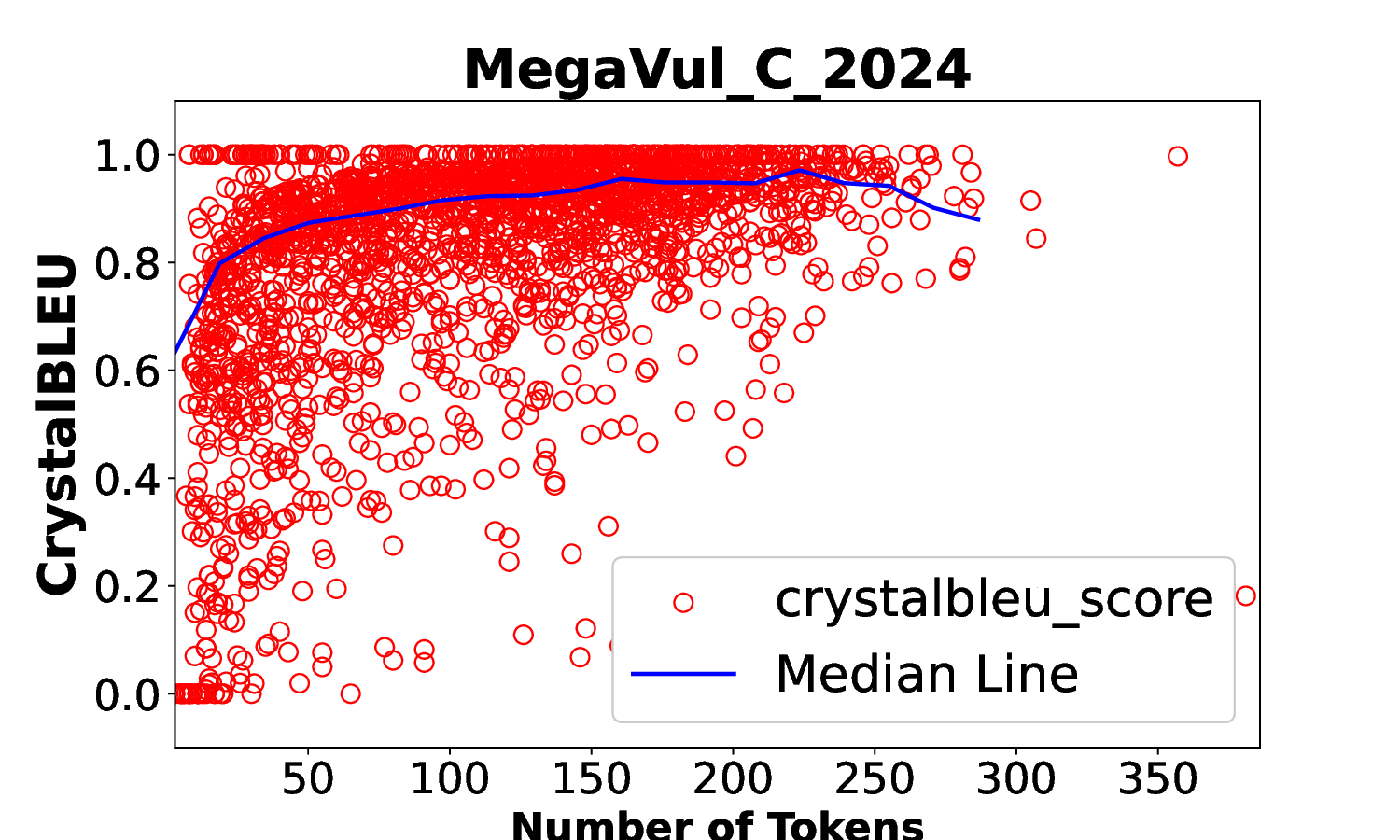}} \\
      
      &  \raisebox{-0.5\height}{\includegraphics[width=0.34\textwidth] {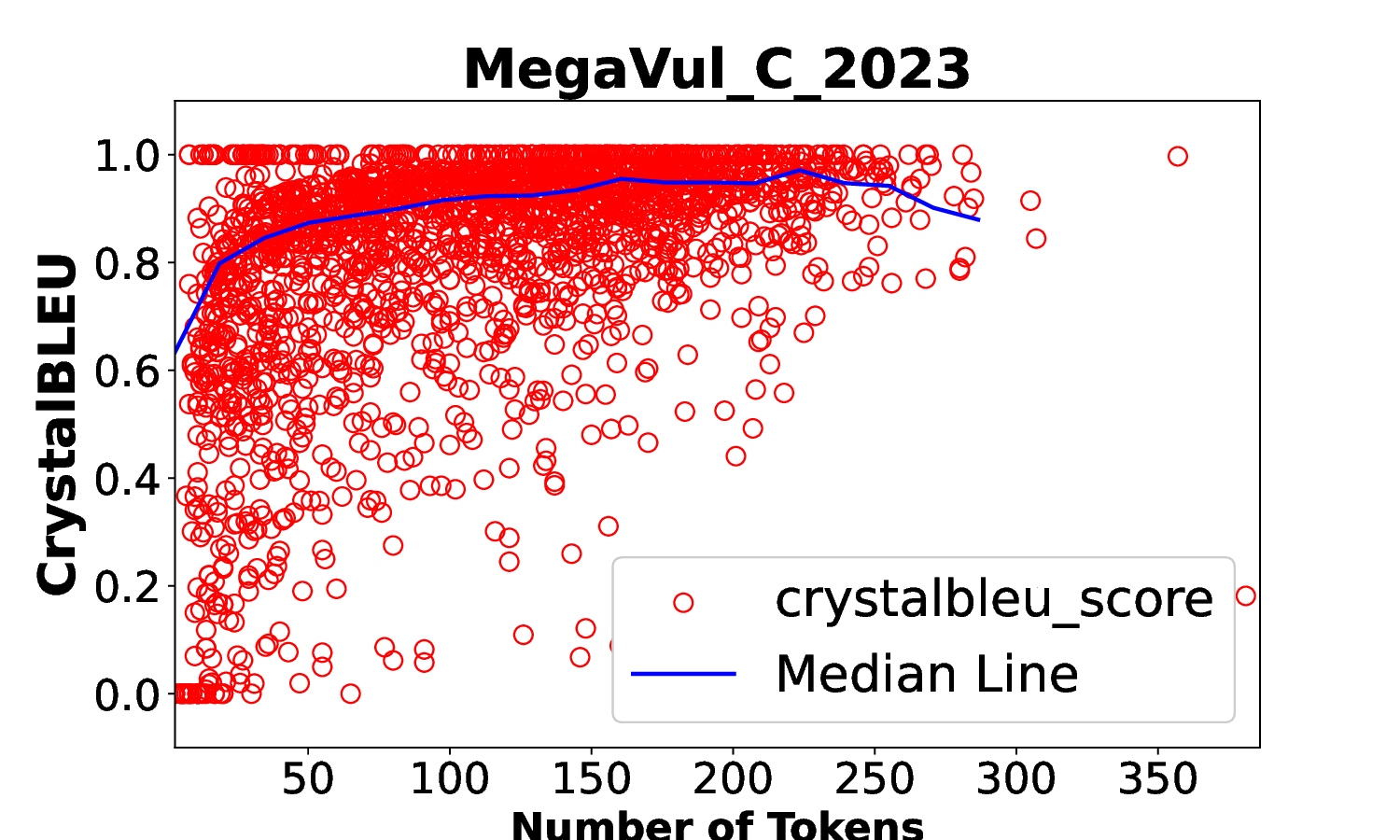}}
      &  \raisebox{-0.5\height}{\includegraphics[width=0.34\textwidth] {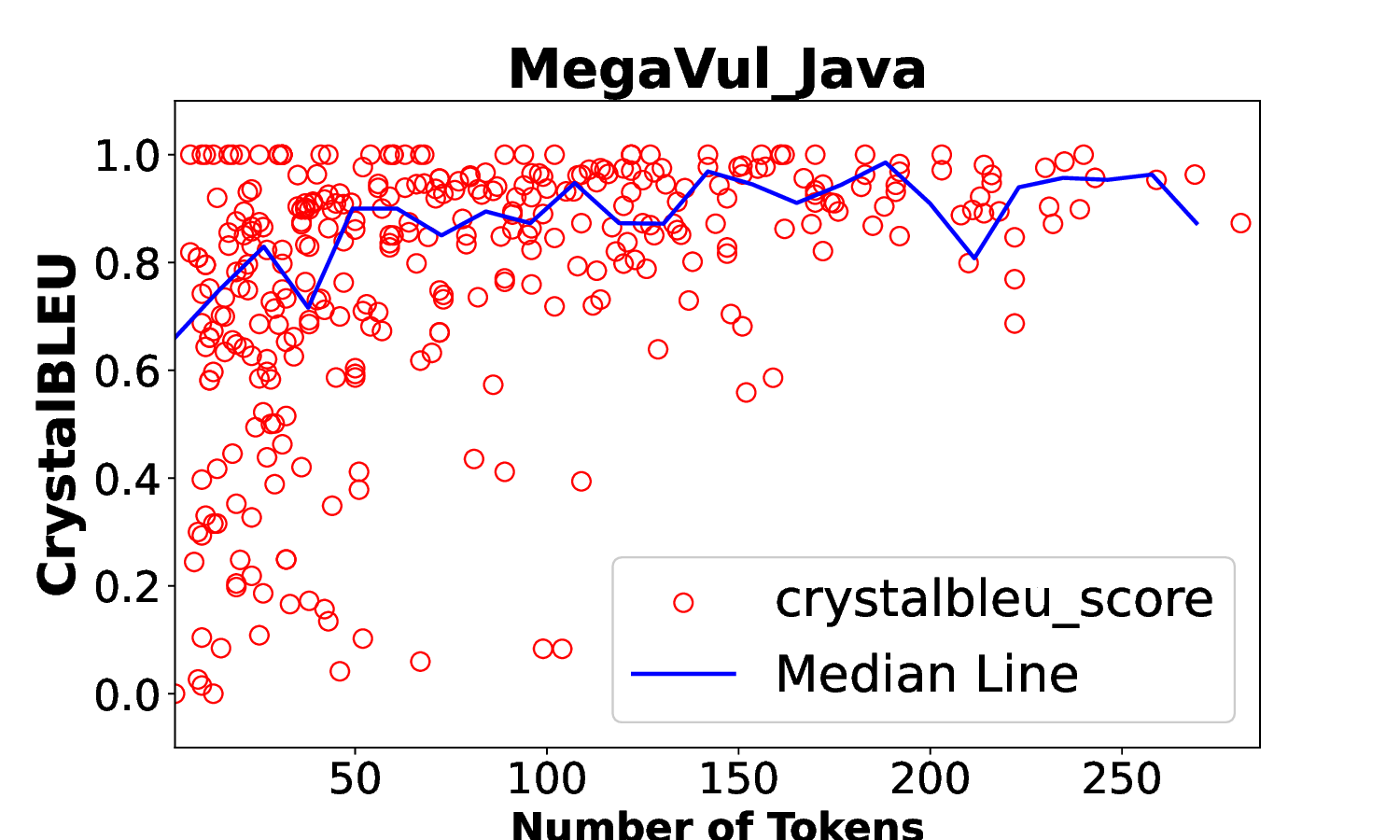}}
      &  \raisebox{-0.5\height}{\includegraphics[width=0.34\textwidth] {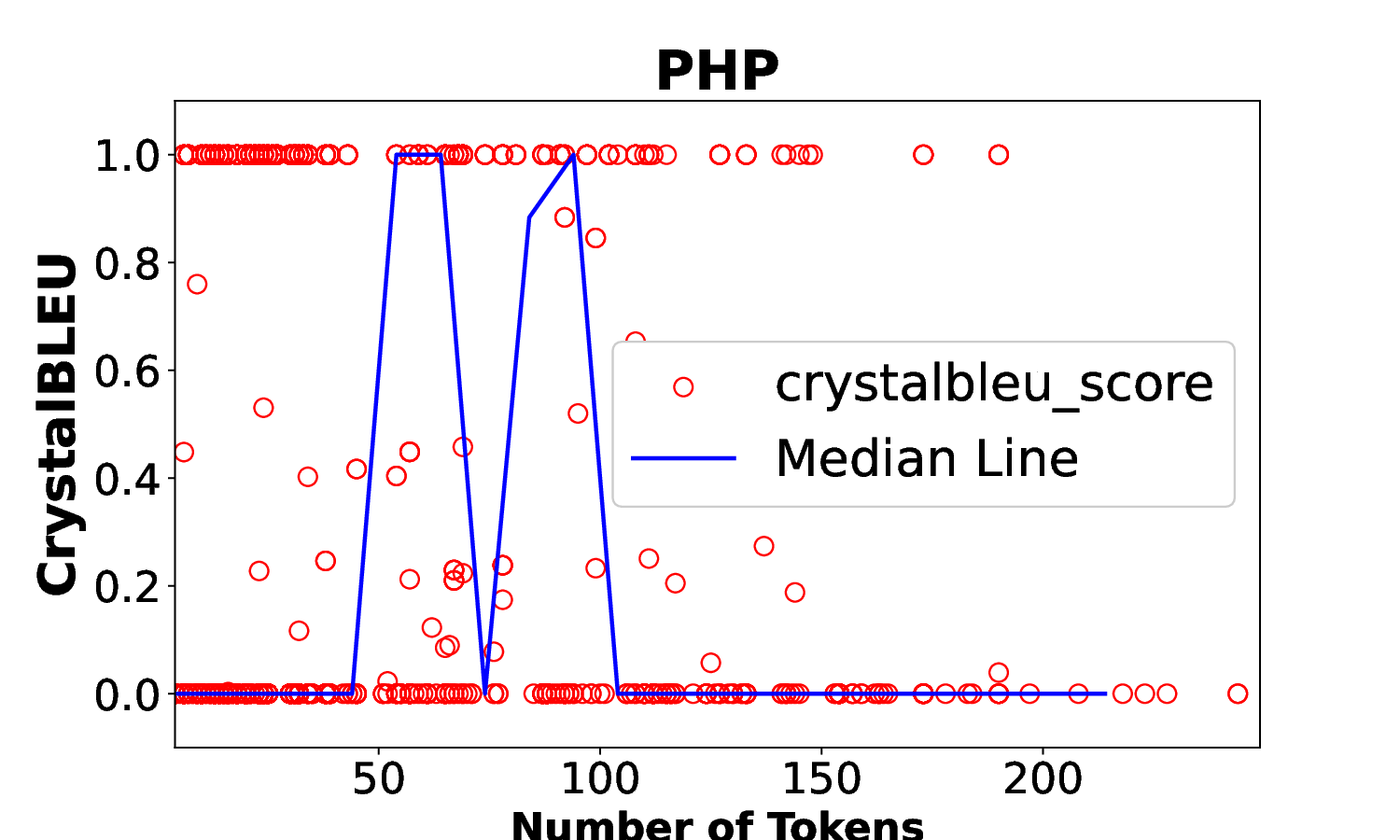}}\\

      &  \raisebox{-0.5\height}{\includegraphics[width=0.34\textwidth] {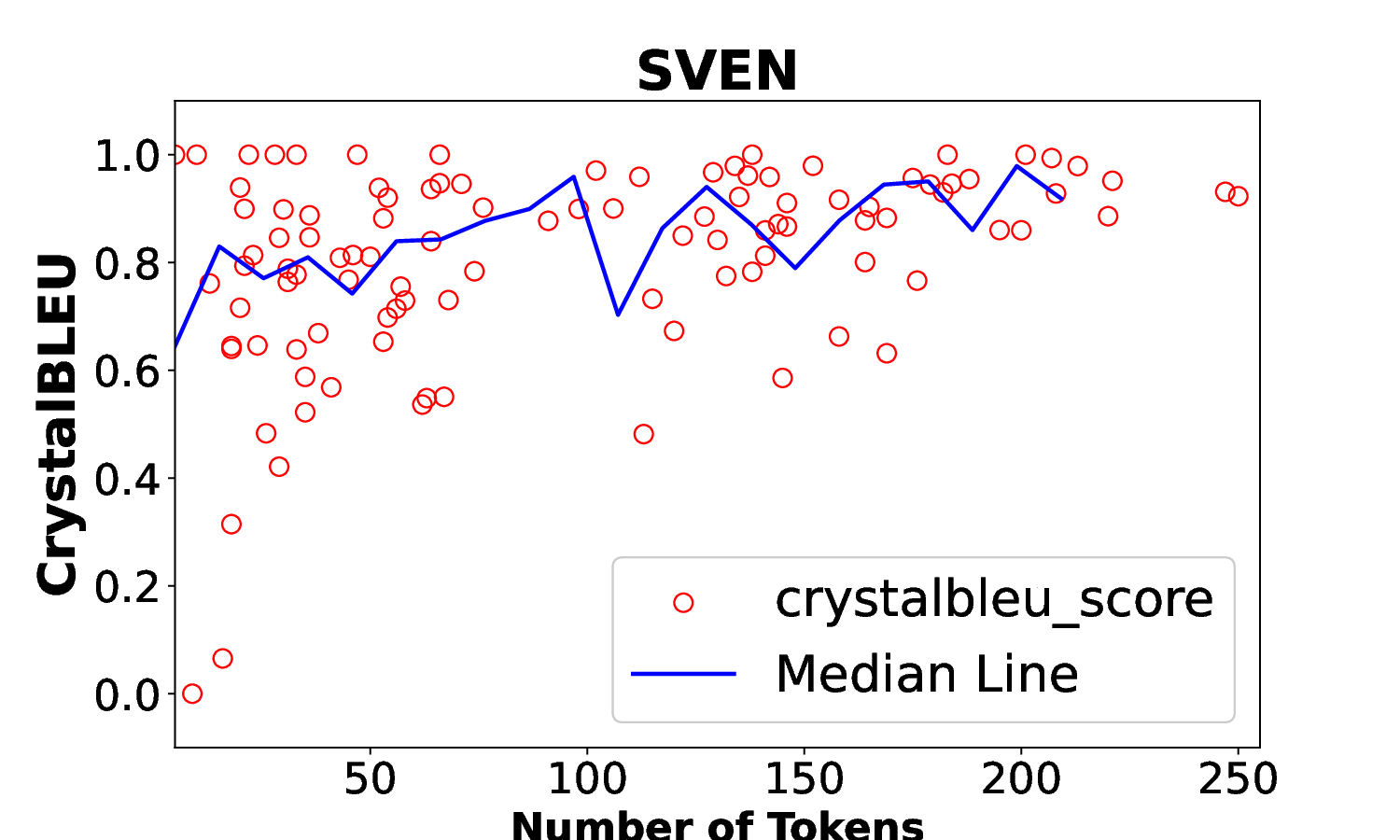}}
      &  \raisebox{-0.5\height}{\includegraphics[width=0.34\textwidth] {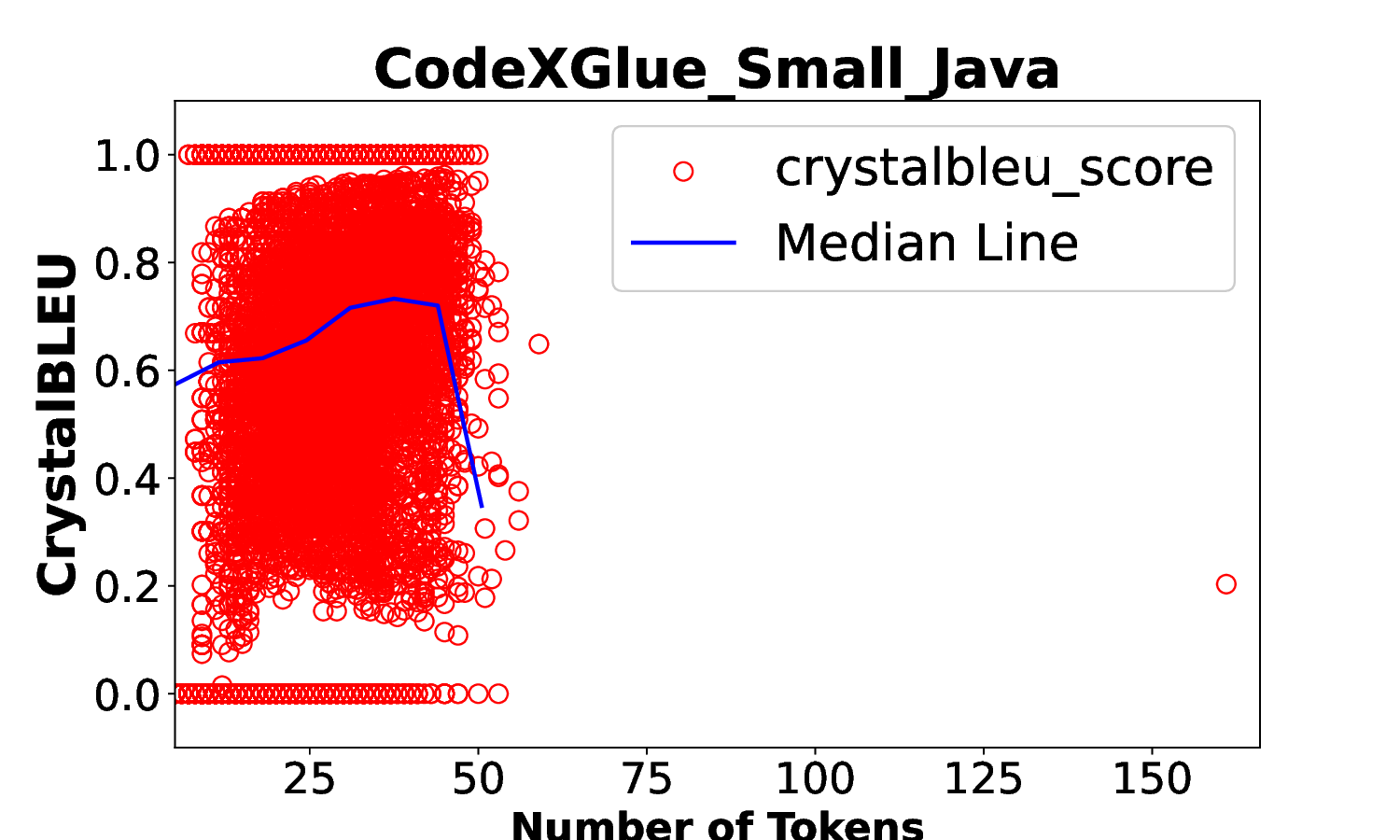}}
      &  \raisebox{-0.5\height}{\includegraphics[width=0.34\textwidth] {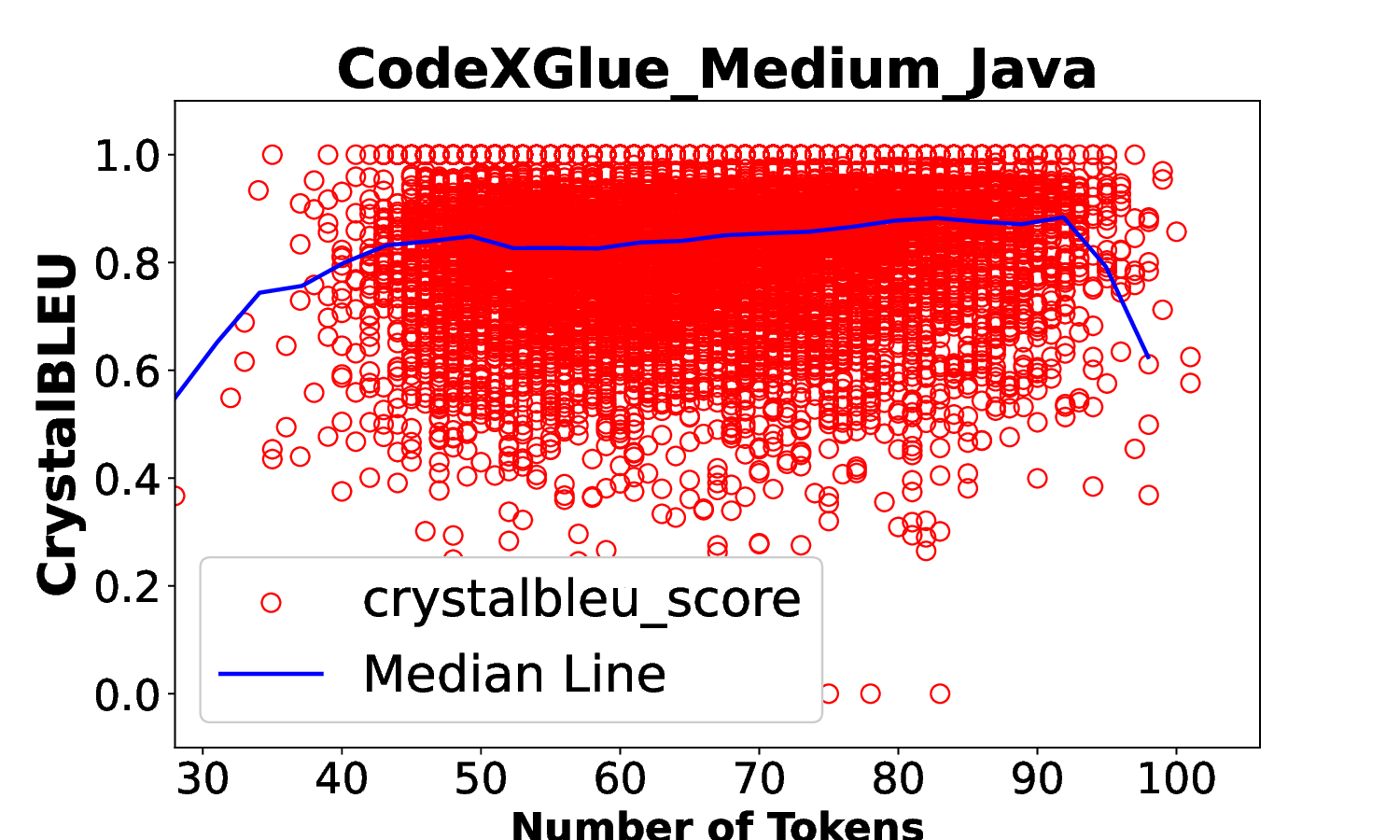}}

    \end{tabularx}
  
  \caption{Correlation Between Patch Length and CrystalBLEU (CodeT5)}
  \label{fig:t5-crystalbleu-graph}
\end{figure}

In addressing RQ3, we examined the impact of patch length on the accuracy of \emph{CodeBERT} and \emph{CodeT5}, using both \emph{CodeBLEU} (Figure~\ref{fig:codebert-codebleu-graph} and Figure~\ref{fig:t5-codebleu-graph}) and \emph{CrystalBLEU} (Figure~\ref{fig:codebert-crystalbleu-graph} and Figure~\ref{fig:t5-crystalbleu-graph}) scores across nine datasets. The results indicate that patch length significantly influences model performance, with each model varying in its ability to handle longer patches.

For \emph{CodeBERT}, accuracy generally declines as patch length increases, as shown by both \emph{CodeBLEU} and \emph{CrystalBLEU} scores. Datasets like BigVul, MegaVul\_C\_2023, and MegaVul\_Java exhibit sharp declines in accuracy for longer patches, highlighting the model's challenges in maintaining coherence over extended sequences. In the PHP dataset, performance fluctuates considerably, demonstrating instability in handling longer patches. However, on the CodeXGlue\_Medium\_Java dataset, \emph{CodeBERT} performs relatively well, maintaining stable scores for longer patches, likely due to the dataset’s consistent structure. In sparse datasets like SVEN, the notably low \emph{CrystalBLEU} scores highlight significant struggles with fragmented contexts, as also observed in \emph{CodeBLEU} evaluations. For the Go dataset, accuracy initially 
drops sharply but later stabilizes for longer patches, which could be 
attributed to the nature of Go code—larger patches might eventually 
capture more relevant contextual information and structural patterns 
that help the model perform better after an initial drop. Additionally, Go's more modular and consistent 
syntax might allow the model to better generalize across longer sequences, 
preventing a further decline in performance.

\emph{CodeT5} exhibits similar variability in performance to \emph{CodeBERT} across most datasets, indicating that both models face challenges in maintaining consistent accuracy with increasing patch lengths. In BigVul, MegaVul\_C\_2023, MegaVul\_C\_2024, and CodeXGlue\_Medium\_Java, both metrics indicate that \emph{CodeT5} generally performs better than \emph{CodeBERT}, demonstrating its relative effectiveness in handling extended sequences. However, maintaining 
accuracy with increasing patch lengths remains a challenge in datasets like PHP, where both models experience a decline in performance. While \emph{CodeT5} demonstrates a marginally better ability to retain accuracy compared to \emph{CodeBERT}, the differences are not substantial. Similarly, the sparse and fragmented data in the Go and SVEN datasets lead to inconsistent accuracy, particularly evident in \emph{CrystalBLEU} scores, highlighting the difficulty of preserving syntactic and semantic integrity in such contexts.

When comparing the two models, \emph{CodeT5} generally demonstrates better performance than \emph{CodeBERT} in handling longer patches, although the differences in accuracy are not substantial across most datasets. Both models exhibit notable challenges with sparse or fragmented datasets, such as Go and SVEN, where longer patch lengths amplify performance issues. The variations observed between \emph{CodeBLEU} and \emph{CrystalBLEU} scores emphasize the difficulties in maintaining both syntactic and semantic integrity in extended patches. This highlights the complexity of generating accurate and contextually coherent patches for vulnerability-focused program repair as patch length increases.

\section{Conclusion}

In this study, we evaluated the effectiveness of two advanced pre-trained language models, CodeBERT and CodeT5, in addressing vulnerability-focused program repair across a varied collection of datasets. Through systematic experimentation, we analyzed their accuracy in patch generation, computational efficiency, and the impact of patch length on performance, providing valuable insights into their potential and limitations for automated vulnerability patching.

Our findings show that CodeT5 outperformed CodeBERT in generating accurate patches, particularly for datasets with complex or diverse vulnerability patterns, such as BigVul and MegaVul. However, CodeBERT exhibited strengths in handling context-limited datasets like Go and PHP, where fragmented or incomplete code posed challenges. 

In terms of computational efficiency, CodeT5 consistently demonstrated faster inference times across nearly all datasets, making it better suited for large-scale applications, such as integration into CI/CD pipelines or vulnerability assessments of extensive codebases. Meanwhile, CodeBERT’s robustness in certain contexts suggests its simpler architecture can be advantageous for specific use cases, such as patching legacy systems or isolated code snippets.

The analysis of patch length revealed challenges for both models in generating accurate and coherent patches for extended sequences. While CodeT5 exhibited a relative advantage in maintaining performance across longer patches, both models experienced declines, emphasizing the need for future advancements in addressing longer contextual dependencies.

These findings underscore the complex interplay between dataset characteristics, model architecture, and task-specific requirements in vulnerability-focused program repair. Although CodeT5 demonstrates superior accuracy and computational efficiency, making it well-suited for large-scale and diverse datasets, its challenges with extended patch lengths reveal the limitations of current pre-trained models in managing long-range dependencies—a critical requirement for security patches in modern software ecosystems that demand nuanced and context-aware modifications. On the other hand, CodeBERT’s performance in context-limited scenarios demonstrates the value of simpler architectures in addressing fragmented or incomplete code, suggesting that no single model can effectively address all aspects of automated vulnerability patching. These findings emphasize the need to align model selection and fine-tuning strategies with the unique characteristics of datasets and use cases, carefully balancing accuracy, efficiency, and adaptability. Future work should focus on hybrid approaches that combine the contextual understanding of advanced models like CodeT5 with the robustness of simpler architectures like CodeBERT, as well as improving dataset quality and diversity to better reflect the challenges of real-world vulnerability repair. Moreover, advancements in handling longer patch lengths, whether through enhanced pre-training strategies or specialized model architectures, are essential to overcome the bottlenecks observed in this study, paving the way for more scalable and effective solutions in automated vulnerability patching.

\begin{credits}

\subsubsection{\ackname}  This research received funding from the European Commission through the Horizon Europe Programme as part of the LAZARUS project (\url{https://lazarus-he.eu/}) (Grant Agreement No. 101070303). The content of this article represents the sole responsibility of the authors and does not necessarily reflect the official views of the European Union.

\subsubsection{\discintname}
The authors have no competing interests to declare that are relevant to the content of this article.
\end{credits}

\newpage
\bibliographystyle{unsrt}
\bibliography{refs}
\end{document}

%% file: tables/datasets.tex
\begin{filecontents*}{data/statistics.csv}
dataset,totalRowsInitial,impactedTokenization,impactedComments,impactedNormalization,totalRowsRemaining,testInstances
Go,1472,551,357,0,921,138
PHP,6696,335,4923,1,6360,954
BigVul,10900,3756,0,0,7144,1072
MegaVul_Java,2433,62,0,75,2296,344
MegaVul_C_2023 ,17975,3147,0,302,14526,2179
MegaVul_C_2024,17975,3147,0,302,14526,2179
SVEN,803,104,0,4,695,104.3
CodeXGlue_Medium_Java,65455,0,0,0,65455,9818.3
CodeXGlue_Small_Java ,58350,0,0,0,58350,8753
\end{filecontents*}
\pgfplotstabletypeset[
    col sep=comma,
    fixed,
    empty cells with={-},
    string replace={---}{\textemdash},
     string replace={_}{\textunderscore}, 
    every head row/.style={
        before row={\toprule},
        after row=\midrule,
    },
    every last row/.style={after row=\bottomrule},
    columns={dataset,totalRowsInitial,impactedTokenization,impactedComments,impactedNormalization,totalRowsRemaining}, 
    columns/dataset/.style={column type=l, column name=Dataset, string type, postproc cell content/.code={%
        \pgfplotsutilstrreplace{_}{\_}{##1}%
        \pgfkeyslet{/pgfplots/table/@cell content}\pgfplotsretval
    },},
    columns/totalRowsInitial/.style={column type=r, column name= $I_{rows}$, precision=4},
    columns/impactedTokenization/.style={column type=r, column name=$R_{tok.}$, precision=4},
    columns/impactedComments/.style={column type=r, column name=$R_{comm.}$, precision=4},
    columns/impactedNormalization/.style={column type=r, column name=$R_{norm.}$, precision=4},columns/totalRowsRemaining/.style={column type=r, column name=$T_{rows}$, precision=4},
]{data/statistics.csv} 

%% file: tables/RQ1_results.tex
\begin{filecontents*}{data/summary_RQ1_RQ2.csv}
Dataset,CodeBLEUBERT,CodeBLEUT5,CrystalBLEUBERT,CrystalBLEUT5,trainTimeBERT,trainTimeT5,testTimeBERT,testTimeT5,totalTimeBERT,totalTimeT5,AvgInferenceTimeBERT,AvgInferenceTimeT5,testInstancesBERT,testInstancesT5
Go,0.76409358,0.649903524,0.655735471,0.52641215,1308.211,2353.508,339.45,173.429,1647.661,2526.937,2.442086331,1.247690647,139,139
PHP,0.735140199,0.692412555,0.462377324,0.372659485,8747.032,16118.246,1405.674,1219.889,10152.706,17338.135,1.47345283,1.278709644,954,954
BigVul,0.625392906,0.814788922,0.414066144,0.616085486,10123.721,18131.706,3043.843,1310.126,13167.564,19441.832,2.839405784,1.222132463,1072,1072
MegaVul_Java,0.273569053,0.831977525,0.04065961,0.774656303,3247.235,5830.893,654.723,488.187,3901.958,6319.08,1.903264535,1.419148256,345,345
MegaVul_C_2023,0.839612819,0.854883959,0.789343064,0.813103693,20323.176,37135.945,4429.201,2653.073,24752.377,39789.018,2.032675998,1.217564479,2179,2179
MegaVul_C_2024,0.839544128,0.85492269,0.789343064,0.813103693,20247.623,36976.005,4422.296,3322.019,24669.919,40298.024,2.029507113,1.524561267,2179,2179
SVEN,0.212708733,0.829734324,0,0.80889947,999.601,1777.1,178.963,159.634,1178.564,1936.734,1.720798077,1.534942308,105,105
CodeXGlue_Medium_Java,0.867446987,0.866683259,0.827685613,0.824006969,89693.675,165263.176,15252.756,3707.061,104946.431,168970.237,1.553550214,0.37757802,9819,9819
CodeXGlue_Small_Java,0.776263965,0.763922383,0.648186931,0.629067284,78671.324,146070.758,4518.48,1658.602,83189.804,147729.36,0.516220724,0.189489546,8753,8753
\end{filecontents*}
\pgfplotstabletypeset[
    col sep=comma,
    fixed,
    empty cells with={-},
    string replace={---}{\textemdash},
    string replace={_}{\textunderscore},
    every head row/.style={
        before row={
	   		\toprule
			\multicolumn{1}{c}{} & 
			\multicolumn{2}{c}{CodeBLEU} &
			\multicolumn{2}{c}{CrystalBLEU} 
			\\
			\cmidrule(r){2-3}
			\cmidrule(r){4-5}
		},
        after row=\midrule,
    },
    every last row/.style={after row=\bottomrule},
    columns={Dataset,CodeBLEUBERT,CodeBLEUT5,CrystalBLEUBERT,CrystalBLEUT5}, 
    columns/Dataset/.style={column type=l, string type, postproc cell content/.code={%
        \pgfplotsutilstrreplace{_}{\_}{##1}%
        \pgfkeyslet{/pgfplots/table/@cell content}\pgfplotsretval
    }},
    columns/CodeBLEUBERT/.style={column type=r,column name=$CodeBERT$, precision=4},
    columns/CodeBLEUT5/.style={column type=r, column name=$CodeT5$, precision=4},
    columns/CrystalBLEUBERT/.style={column type=r,column name=$CodeBERT$, precision=4},
    columns/CrystalBLEUT5/.style={column type=r, column name=$CodeT5$, precision=4},
]{data/summary_RQ1_RQ2.csv}

%% file: tables/RQ2_results.tex
\pgfplotstabletypeset[
    col sep=comma,
    fixed,
    empty cells with={-},
    string replace={---}{\textemdash},
    string replace={_}{\textunderscore},
    every head row/.style={
        before row={\toprule},
        after row=\midrule,
    },
    every last row/.style={after row=\bottomrule},
    columns={Dataset,AvgInferenceTimeBERT,AvgInferenceTimeT5,testInstancesBERT}, 
    columns/Dataset/.style={column type=l, string type, postproc cell content/.code={%
        \pgfplotsutilstrreplace{_}{\_}{##1}%
        \pgfkeyslet{/pgfplots/table/@cell content}\pgfplotsretval
    }},
    columns/AvgInferenceTimeBERT/.style={column type=r,column name=$CodeBERT$, precision=4},
    columns/AvgInferenceTimeT5/.style={column type=r, column name=$CodeT5$, precision=4},
    columns/testInstancesBERT/.style={column type=r,column name=$Test Instances$, precision=4},
]{data/summary_RQ1_RQ2.csv}